\documentclass[preprint,showpacs,preprintnumbers, amsmath,amssymb,superscriptaddress,nofootinbib]{revtex4-1}
\usepackage{graphicx,color}
\usepackage{amsmath,amssymb}
\usepackage{url}
\usepackage{epstopdf}

\newcommand{\hs}{\hspace*{0.5cm}}

\newcommand{\be}{\begin{equation}}
\newcommand{\ee}{\end{equation}}
\newcommand{\bea}{\begin{eqnarray}}
\newcommand{\eea}{\end{eqnarray}}

\newcommand{\crn}{\nonumber \\}

\newcommand{\bc}{\begin{center}}
\newcommand{\ec}{\end{center}}

\newcommand {\ba}{\begin{array}}
\newcommand {\ea}{\end{array}}
\newcommand{\ben}{\begin{enumerate}}
\newcommand{\een}{\end{enumerate}}

\usepackage{epsfig,graphicx}
\usepackage{bm}
\usepackage{dcolumn}

\begin{document}

\title{ Heavy singly charged Higgs bosons and inverse seesaw neutrinos  as origins of large  $(g-2)_{e,\mu}$ in  two Higgs doublet models}

\author{L. T. Hue}
\email{lethohue@vlu.edu.vn}
\affiliation{Subatomic Physics Research Group, Science and Technology Advanced Institute, Van Lang University, Ho Chi Minh City 70000, Vietnam}
\affiliation{Faculty of Applied Technology, School of Engineering and Technology, Van Lang University, Ho Chi Minh City 70000, Vietnam}

\author{A. E. C\'arcamo Hern\'andez}
\email{antonio.carcamo@usm.cl}
\affiliation{Universidad T\'ecnica Federico Santa Mar\'{\i}a, Casilla 110-V, Valpara%
	\'{\i}so, Chile}
\affiliation{Centro Cient\'{\i}fico-Tecnol\'ogico de Valpara\'{\i}so, Casilla 110-V,
	Valpara\'{\i}so, Chile}
\affiliation{Millennium Institute for Subatomic Physics at High-Energy Frontier
	(SAPHIR), Fern\'andez Concha 700, Santiago, Chile}

\author{ H. N. Long } 
\email{hnlong@iop.vast.vn}
\affiliation{Institute of Physics,   Vietnam Academy of Science and Technology, 10 Dao Tan, Ba Dinh,  Hanoi 10000, Vietnam}

\author{T.~T.~Hong \footnote{Corresponding author}}
\email{tthong@agu.edu.vn}
\affiliation{An Giang University, VNU - HCM, Ung Van Khiem Street,
	Long Xuyen, An Giang 88000, Vietnam}


\begin{abstract}
We show that simple extensions of two Higgs doublet models consisting of new heavy neutrinos and a singly charged Higgs boson singlet can successfully explain the experimental data on muon and electron anomalous magnetic moments  thanks to large chirally-enhanced one-loop level contributions. These contributions arise from the large couplings of inverse seesaw neutrinos with singly charged Higgs bosons and right-handed charged leptons. The regions of parameter space satisfying the experimental data on $(g-2)_{e,\mu}$ anomalies allow heavy singly charged Higgs boson masses above the TeV scale, provided that heavy neutrino masses are above few hundred GeV, the non-unitary part of the active neutrino mixing matrix must be large enough, two singly charged Higgs bosons are non degenerate, and the mixing between singly charged Higgs bosons must be non-zero.
\end{abstract}
\maketitle
\section{Introduction}
\label{sec:intro}
\allowdisplaybreaks
The latest experimental measurement for the anomalous magnetic moment (AMM) of the muon $a_{\mu}\equiv (g-2)_{\mu}/2$ has been reported from   
Fermilab~\cite{Abi:2021gix} and is in agreement with the previous experimental result measured by Brookhaven National Laboratory (BNL) E82~\cite{Muong-2:2006rrc}. A combination of these  results in the new average value of $a^{\mathrm{exp}}_{\mu}=116592061(41)\times 10^{-11}$, which leads to the improved standard deviation of 4.2 $\sigma$ from the Standard Model  (SM) prediction, namely 
\begin{equation}\label{eq_damu}
	\Delta a^{\mathrm{NP}}_{\mu}\equiv  a^{\mathrm{exp}}_{\mu} -a^{\mathrm{SM}}_{\mu} =\left(2.51\pm 0.59 \right) \times 10^{-9},
\end{equation} 
 where  $a^{\mathrm{SM}}_{\mu}= 116591810(43)\times 10^{-11}$ is the  SM prediction \cite{Aoyama:2020ynm} combined from various different contributions \cite{Davier:2010nc, Davier:2017zfy, Keshavarzi:2018mgv, Colangelo:2018mtw, Hoferichter:2019mqg, Davier:2019can, Keshavarzi:2019abf, Kurz:2014wya, Melnikov:2003xd, Masjuan:2017tvw, Colangelo:2017fiz, Hoferichter:2018kwz, Gerardin:2019vio, Bijnens:2019ghy, Colangelo:2019uex, Colangelo:2014qya, Blum:2019ugy, Aoyama:2012wk, Aoyama:2019ryr, Czarnecki:2002nt, Gnendiger:2013pva}.   On the other hand, the recent experimental $a_e$ data  was  reported from different groups~\cite{Hanneke:2008tm, Parker:2018vye, Morel:2020dww}, leading to the latest deviation  between experiment and the SM prediction \cite{ Aoyama:2012wj, Aoyama:2012wk, Laporta:2017okg, Aoyama:2017uqe,  Terazawa:2018pdc, Volkov:2019phy, Gerardin:2020gpp} as follows: 
\begin{equation}\label{eq_dae}
	\Delta a^{\mathrm{NP}}_{e}\equiv  a^{\mathrm{exp}}_{e} -a^{\mathrm{SM}}_{e} = \left( 4.8\pm 3.0\right) \times 10^{-13}.
\end{equation}  
Many Beyond Standard Model (BSM) theories
have been constructed to explain the
$(g-2)_{\mu,e}$ anomalies. Such theories rely  
on the inclusion of vector-like lepton multiplets \cite{Dermisek:2013gta, Crivellin:2018qmi, Escribano:2021css,Hernandez:2021tii, Crivellin:2021rbq, Dermisek:2021ajd, Chun:2020uzw,Frank:2020smf,  Endo:2020tkb, Chen:2020tfr, Hati:2020fzp, Ferreira:2021gke, Borah:2021khc, Bharadwaj:2021tgp, DeJesus:2020yqx, Cogollo:2020nrc, Saez:2021qta}, leptoquarks \cite{Bigaran:2020jil, Crivellin:2020tsz, Zhang:2021dgl, Keung:2021rps}, both neutral and charged  Higss bosons as $SU(2)_L$ singlets \cite{Mondal:2021vou, Kang:2021jmi}.  Some two Higgs doublet models  (THDM) can provide sizeable 
two-loop contributions to $\Delta a_{\mu}$ arising from new $SU(2)_L$ Higgs doublets ~\cite{CarcamoHernandez:2021iat, Li:2020dbg, DelleRose:2020oaa, Botella:2020xzf, Han:2021gfu, Chen:2021jok, Han:2018znu, Rose:2021cav}, where some of them require rather light masses of  new neutral and/or charged Higgs bosons at few hundred GeV.  

There are many types of  different versions of THDM, such as for instance: type I, type II, type X and type Y, which are discussed in Ref.~\cite{Aoki:2009ha}, where collider phenomenology will give rise to different physical results. They are distinguished among themselves  by the ways in which the  two different $SU(2)_L$ Higgs doublets $\Phi_{1,2}$ generate the masses of up and down quarks as well as of charged leptons. These models arise from   suitable choices of $Z_2$ charge assignments.  The Yukawa couplings of Higgs bosons  with heavy fermions will be constrained by the perturbative limits, for example $0.4\leq \tan\beta \leq 91$ corresponding to the definitions  of Yukawa couplings of the top and bottom quarks $\left| Y_t\right|=(\sqrt{2}/v)m_t\cot\beta<\sqrt{\pi}$ and $\left| Y_b\right|=(\sqrt{2}/v)m_b\tan \beta<\sqrt{\pi}$, where $\tan\beta$ is the ratio between  
the two vacuum expectations values (vevs) of the two neutral Higgs components.  Discussions on the original THDM  have  shown that a new $SU(2)_L$ Higgs doublet can be used to accommodate the $(g-2)_{e,\mu}$ anomalies  in a rather narrow allowed region of parameter space at the price of imposing 
dangerous requirements that can be experimentally tested in the near future \cite{Li:2020dbg, Jueid:2021avn}. The conclusion is also true for the Minimal Supersymmetric Model (MSSM)  \cite{Badziak:2019gaf, Li:2021koa, Athron:2021iuf}.  Another solution for MSSM requires large SUSY threshold corrections enough to change both sizes and signs of  the SUSY electron and muon Yukawa couplings \cite{Endo:2019bcj}, hence allow regions satisfying both $(g-2)_{e,\mu}$ anomalies with large TeV scale of slepton masses, but  rather large $t_{\beta}\ge70$ is needed.  As usual solution, there is a variety of  extensions of  
THDM that successfully explain and accommodate 
the $(g-2)_{e,\mu}$ anomalies  by adding new vector-like  fermions as $SU(2)_L$ multiplets  \cite{Chun:2020uzw,Frank:2020smf,  Hernandez:2021tii, Chen:2020tfr, Dermisek:2021ajd, Ferreira:2021gke} as well as  $SU(2)_L$ singlets of neutral and charged Higss bosons \cite{CarcamoHernandez:2019xkb, CarcamoHernandez:2019cbd, Mondal:2021vou, CarcamoHernandez:2021iat, Hernandez:2021xet, Adhikari:2021yvx}. Only few THDM models with scalar singlets  and fermions can successfully explain the experimental data of both muon and electron anomalous magnetic moments~\cite{Chen:2020tfr, Dutta:2020scq, Hernandez:2021xet}. Our work here will introduce a more general solution that a wide class of the THDM by adding heavy right handed (RH) neutrino singlets 
and one singly electrically charged Higgs boson can give sizeable 
one-loop contributions enough to explain the experimental ranges of values of both muon and electron anomalous magnetic moments in a wide region of parameter space.

 This paper is organized as follows.  In Sec.~\ref{sec_2HDM}, the THDM extension with  
  new RH neutrinos and singly charged Higgs boson (THDM$N_Rh^\pm$)  will be introduced, where we pay attention to the leptons, gauge bosons, and Higgs sectors, giving all physical states as well the couplings that may give large one-loop contributions to AMM. In Sec. \ref{eq_Deltaamu}, the simple inverse seesaw (ISS) and  analytic formulas for one-loop contributions to AMM are constructed. Numerical discussion will be shown in details.  Our main results will be collected in Sec.~\ref{eq_conclusion}. 

\section{ \label{sec_2HDM} Review of the THDM$N_Rh^\pm$}
In this section we will focus on the analysis of the lepton sector of the THDM$N_Rh^\pm$. For details of the quark sector of different types of the THDM, the reader is referred to Refs.~\cite{Davidson:2005cw, Herrero-Garcia:2017xdu}. The leptonic, quark and scalar spectrum with their assignments under the $SU(2)_L\times U(1)_Y$ gauge group are given by ~\cite{Davidson:2005cw, Herrero-Garcia:2017xdu}:
\begin{align}
L_{a}&=	\begin{pmatrix}
		\nu_{aL} \\ e_{aL} 
	\end{pmatrix}
	\sim (2, -1); \;\quad Q_{aL}= \begin{pmatrix}
		u_{aL}\\ d_{aL} 
	\end{pmatrix} 
	\sim \left( 2, \dfrac{1}{3}\right),
	\crn 	e_{aR} &\sim (1,-2);\quad   u_{aR} \sim \left(1,\dfrac{4}{3}\right); \quad  d_{aR} \sim  \left(1, -\dfrac{2}{3}\right); \quad  N_{IR}\sim (1,0), \; a=1,2,3, \; I=1,2,...,6;
\crn & \Phi_i=	\begin{pmatrix}
	\phi^+_i\\\phi^0_i 
\end{pmatrix}
\sim (2, 1), \;  \langle  \Phi_i \rangle =	\begin{pmatrix}
0\\
\frac{v_i}{\sqrt{2}} 
\end{pmatrix},\; i=1,2;
\crn h^\pm&\sim (1,\pm2).  
\end{align}
Note that we use the old convention for the defining the hypercharge via the original Gell-Mann- Nishijima relation:
\begin{equation}
	Q=T_3+\frac1{2}Y.  
\end{equation}
Here we add six new neutrino singlets to the leptonic spectrum of the model in order to implement 
the standard ISS mechanism.  Furthermore, the model scalar sector is augmented by the inclusion of
singly charged Higgs bosons, which allow to generate the new Yukawa interaction $\overline{(N_{IR})^c}e_{aR}h^+$ thus resulting in sizeable
one-loop contributions to AMM. 

 One of the most important parameters introduced in the THDM is 
\begin{equation}\label{eq_tb}
	t_\beta \equiv \tan\beta=\dfrac{v_2}{v_1},   \; s_\beta\equiv \sin \beta, \; c_\beta\equiv \cos \beta. 
\end{equation}
We will also follow the notations $s_x\equiv \sin x$,   $c_x\equiv \cos x$, and $t_x=s_x/c_x$ for any parameter $x$.  Because the Yukawa couplings of up type quarks and $\Phi_2$ are fixed, we always have a lower bound $t_{\beta}\geq 0.4$. In the THDM type I, all fermions couple with only $\Phi_2$, hence no upper bounds for $t_{\beta}$ arise.  In the THDM type II and Y, all down type quarks and charged leptons couple with $\Phi_1$, hence $t_{\beta}\le91$. In the type-X, the Yukawa coupling of tau with $\Phi_1$ gives $t_{\beta}\le 348$.

The THDM type-II and type-X where lepton doublets couple with only $\Phi_1$ are called THDM type-A. 
The corresponding lepton Yukawa terms are:
\begin{align}
\label{eq_ylepton1}
-\mathcal{L}^\text{Y}_{\ell} = & Y^e_{ab} \overline{e_{aR}} \Phi_1^{\dagger}L_{b}  + Y^N_{k,Ia} \overline{N_{IR}}\left(i\sigma_{2} \Phi_k \right) ^TL_{a} + \frac{1}{2}M_{N,IJ} \overline{N_{IR}}(N_{JR})^c 
\crn &  + Y^{h}_{Ia}\overline{(N_{IR})^c} e_{aR} h^+ +\mathrm{h.c.}, 	
\end{align}
where $a,b=1,2,3$ are family indices; $k=1,2$;  and $I=1,2,3,...6$ are the number of new neutral lepton singlets. A rough 
condition for the  pertubative limit of  $Y^{h}$ is  $|Y^{h}_{Ia}|<\sqrt{4\pi}$, and should be smaller for trusty values \cite{Allwicher:2021rtd}.   The $Z_2$ charges of $N_{IR}$ and $h^\pm$ are always chosen to guarantee the $Z_2$ symmetry of Lagrangian \eqref{eq_ylepton1}. Using the $Z_2$ charge assignments discussed in Ref.~\cite{Aoki:2009ha}, two Higgs triplets $\Phi_{1,2}$  have different $Z_2$ charges, therefore $N_{IR}$ couple with only one of them. In addition, we can consider the inclusion of new $Z_3$ discrete symmetry in the model in order to forbid the quartic scalar interaction 
$\left[ \left( \Phi_1^\dagger\Phi_2\right)^2 +\mathrm{H.c.}  \right]$ that generates  active neutrino masses at the one-loop level \cite{Ma:2006km}. This guarantees that the Yukawa couplings $Y^h_{Ia}$  do not have suppressed constraints arising from the neutrino oscillation experimental data.
The detailed assignments of the $Z_2$ and $Z_3$ charges are shown in table~\ref{t_Z23}.
\begin{table}[ht]
	\caption{
		$Z_{2}\times Z_{3}$ charge assignments of the model type-A for 
		the case  I (II) with 
		heavy neutrinos couplings with $\Phi_1$ ($\Phi_2$), $w=e^{i2\pi /3}$ \label{t_Z23}}
	\begin{tabular}{c|cccccccccc}
		&$Q_{aL}$&$u_{aR}$& $d_{aR}$& $L_a$& $e_{aR}$& $N_{1,2,3R}$& $N_{4,5,6R}$& $\Phi_1$&$\Phi_2$& $ h^+$\\
		\hline 
		$Z_2$&-&-&-	&1&1& 1(-1)&-1(1) &1&-1&-1(1)\\
		\hline 
		$Z_3$&$w$&	$w^2$& $w^2$ &$w$&$w^2$  &1& 
		$w$&$w^2$& $w^2$& $w$ \\	
	\end{tabular}	
\end{table} 
The $Z_3$ charges of leptons are chosen for the implementation of the ISS mechanism in the total neutrino mass matrix, namely: i) the charged lepton mass term always consists of $\Phi_1$, ii)  only $N_{aR}$ ($a=1,2,3$)  generate  non-zero Dirac mass term of neutrinos, iii) and only $N_{(a+3)R}$ couple with $h^{\pm}$. The $Z_2$ charges of  quarks depend on which types of  well-known  THDMs, hence we do not list here. The $Z_3$ charges of all quarks have the same values as those of lepton doublets and charged lepton singlets. This guarantees that the quark Yukawa terms are consistent with those previously discussed. 

Different from the well-known THDM, the model under considerations have new kind of Yukawa couplings with singly charged Higgs boson $h^\pm$ and right-handed neutrinos $N_{I}$. In the SM, the vacuum stability around the Plank scale requires the quartic coupling in the Higgs potential close to zero, and the upper bound of the SM Higgs boson mass is very close to the experimental value, $M_h<126$ GeV \cite{Degrassi:2012ry}. This problem can be solved in the model by adding new scalars such as in the THDM \cite{Jangid:2020qgo}, and even with the models by adding neutrino gauge singlets \cite{Jangid:2020dqh, Coriano:2015sea, DelleRose:2015bms}. The reason is that additional scalar couplings to Higgs doublets generating SM-like Higgs boson mass give positive one-loop contributions to the $\beta$-functions of the quartic couplings, hence the negative values of these couplings implying the vacuum instability will be relaxed at higher energy scale. This consequence still hold with the appearance of large Yukawa couplings of new fermions with Higgs doublets,  which give negative one-loop contributions to the $\beta$-functions of the quartic couplings. On the other hand, the upper bounds of these Yukawa couplings are required besides their bounds from the perturbative and  unitarity limits, previously discussed for THDM in refs. \cite{Bandyopadhyay:2020djh}, for example. This situation also applies to the THDM$N_Rh^\pm$ model under consideration. On the other hand, the appearance of the singly charged Higgs boson $h^\pm$ resulting on the Yukawa coupling matrix $Y^h$, which is irrelevant with any Higgs doublets. Hence $Y^h$ is not constrained by the requirement of  vacuum stability. Furthermore, the quartic couplings of $h^\pm$ with Higgs doublets in the Higgs potential will result in the higher energy scale satisfying the vacuum stability.

In the THDM type-I and type-Y, where charged lepton  couple with $\Phi_2$, the first term of Eq. \eqref{eq_ylepton1} should replace $\Phi_1$ with $\Phi_2$. We call them the THDM type-B and will discuss later that the qualitative results for AMM do not change. This model type including the model introduced in Ref. \cite{Chen:2020tfr}.

We also emphasize that the models under consideration including the Yukawa part discussed on Ref.  \cite{Mondal:2021vou} focusing on one particular case of $\Phi_k=\Phi_2$.  As we will show that in our numerical analysis,  the regions of the parameter space predicting $\Delta a_{e,\mu}$ consistent with the experimental data, favor small values of $t_{\beta}$, including the range $t_{\beta} \in [0.4,\; 10]$ which was not mentioned in Ref.~\cite{Mondal:2021vou}. 

Defining  $\nu'_L=(\nu_1, \nu_2, \nu_3)_L^T$ and $N_R=(N_1, N_2,...,N_6)_R^T$, we find the following leptonic mass terms: 
\begin{align}
	-\mathcal{L}^\text{mass}_{\mathrm{lepton}}&=  \frac{Y^e_{ab}v_1}{\sqrt{2}} \overline{e'_{aR}}e'_{bL}  + \frac{1}{2} \begin{pmatrix}
		\overline{(\nu'_L)^c}& \overline{N_R}  
	\end{pmatrix}
	\mathcal{M}^{\nu} 
	\begin{pmatrix}
\nu'_L		\\
(N_R)^c		
	\end{pmatrix}
	   +\mathrm{h.c.},  \quad 	\mathcal{M}^{\nu}  = \begin{pmatrix}
	 0_{3\times 3}& M^T_D \\
	 M_D&  M_N
	   \end{pmatrix}
	 	   \; ,\label{eq_mlepton1}
\end{align}
where 
\begin{equation}\label{eq_MD}
(M_D)_{Ia}\equiv  M_{D,Ia}= \sum_{k=1}^2\frac{Y^N_{k, Ia} v_k}{\sqrt{2}}	
\end{equation}
are the components of a $6\times 3$ Dirac neutrino mass matrix and $M_N$ is a $6\times 6$ symmetric Majorana mass matrix. Here we choose basis where the charged lepton mass matrix is diagonal, which implies $Y^e_{ab}=\delta_{ab}\sqrt{2}m_{e_a}/v_1$, i.e., the flavor and mass states of charged leptons are the same.  The total unitary mixing matrix is defined as 
\begin{align}
	\label{eq_Unu}
	U^{\nu T} 	\mathcal{M}^{\nu} U^{\nu } &= \hat{\mathcal{M}}^{\nu}=\mathrm{diag}(m_{n_1},\;m_{n_2},\;...,m_{n_{9}}) \equiv \mathrm{diag}(\hat{m}_{\nu},\; \hat{M}_{N}),
	\crn 	\begin{pmatrix}
		\nu'_L		\\
		(N_R)^c		
	\end{pmatrix} &= U^{\nu} n_{L},\; 
\begin{pmatrix}
(\nu'_L)^c		\\
N_R		
\end{pmatrix} = U^{\nu*} n_{R}= U^{\nu*} (n_{L})^c,\; 
\end{align} 
where $ n_{L,R}=(n_{1},n_{2},..., n_{9})_{L,R}$ are the Majorana neutrino mass eigenstates satisfying $n_{iL,R}=(n_{iR,L})^c$, and the four-component forms are  $n_i=( n_{iL}, n_{iR})^T$. 

Given that we are working in the basis where the charged lepton mass matrix is diagonal, the leptonic mixing entirely arises from the neutrino sector. This implies that, in order to successfully reproduce the neutrino oscillation experimental data, the neutrino mixing matrix is parameterised in the following form \cite{Ibarra:2010xw}: 
\begin{equation} 
U^{\nu}= \left(
\begin{array}{cc}
	\left(	I_3-\frac{1}{2}RR^{\dagger} \right) U_{\mathrm{PMNS}} & RV \\
	-R^\dagger U_{\mathrm{PMNS}} & \left(I_6 -\frac{1}{2}R^{\dagger} R\right)V \\
\end{array}
\right)  +\mathcal{O}(R^3), 
\label{eq_Unu0}	
\end{equation}
where $V$ is a $6\times 6$ unitary matrix; $R$ is a $3\times 6$ matrix satisfying $|R_{aI}|<1$ for all $a=1,2,3$, and $I=1,2,...,6$. The $3\times 3$ unitary matrix  $U_{\mathrm{PMNS}}$ is the Pontecorvo-Maki-Nakagawa-Sakata (PMNS) matrix \cite{ParticleDataGroup:2020ssz}.    The condition $|R_{aI}|<1$ needed to guarantee that the $U^{\nu}$ as a  power series of $R$ given in Ref. \cite{Ibarra:2010xw} is convergent and the approximation up to the order $R^2$ results in  seesaw relations well-known in literature. Namely,  the total neutrino mass matrix given in Eq. \eqref{eq_mlepton1} will result in three light masses for active neutrinos and new heavy neutrinos, $m_{n_{1,2,3}}\ll m_{n_{4,5,\dots}}$. This form of $U^{\nu}$ is consistent with many other well-known parameterizations \cite{Schechter:1981cv, Korner:1992zk, Grimus:2000vj}.

The successful implementation of the ISS mechanism requires to construct the Dirac and Majorana mass matrices in terms of $3\times 3$ submatrices as follows \cite{Arganda:2014dta, Ibarra:2010xw} 
 \begin{align}
 	M^T_D= (m^T_D,\; 0_{3\times3}), \hs M_N=\left(
 	\begin{array}{cc}
 		0_{3\times3}& M_R \\
 		M^T_R & \mu_X \\
 	\end{array}
 	\right), 
 	\label{repara}
 \end{align}
 where $0_{3\times3}$ is  the $3\times3$ null matrix. 
 Using a new notation  
 $M=M_R\mu_X^{-1}M_R^T$,   we have the following ISS relations: 
  \begin{align}
  	R&= M^{\dagger}_D{M^*_N}^{-1}  = \left(-m_D^{\dagger}M^{*-1},\hs m^\dagger_D\left(M^\dagger_R\right)^{-1} \right),
  	 \crn U^*_{\mathrm{PMNS} } \hat{m}_{\nu} U^\dagger_{\mathrm{PMNS} } &= m_{\nu}=-M^T_DM_N^{-1}M_D =  m_D^T \left( M_R^T\right)^{-1}\mu_XM_R^{-1}m_D,
  	\crn V^*\hat{M}_NV^{\dagger} &\simeq M_{N} + \frac{1}{2}R^TR^*M_N +\frac{1}{2}M_N R^{\dagger}R.  
  	\label{eq_RISS}
  \end{align}
  Because  $m_D$ is parameterized in terms of many free parameters, hence it is enough to choose $\mu_X=\mu_0 I_3$ with $\mu_0>0$, $M_R=\hat{M}_R= M_0I_3$. We choose a simple form $m_D= \frac{M_0}{\sqrt{\mu_0}}\sqrt{\hat{m}_\nu}U^{\dagger}_{\mathrm{PMNS}}$ \cite{Casas:2001sr, Arganda:2014dta, Ibarra:2010xw}. 
 The ISS condition $|\hat{m}_{\nu}|\ll |\mu_X|\ll |m_D|\ll M_0$  gives $\frac{\sqrt{\mu_0  \hat{m}_{\nu}}}{M_0}\simeq0$ and 
 \begin{align}
\hat{M}_N= \left(\begin{matrix}
 	\hat{M}_R	& 0_{3\times 3} \\ 
 	0_{3\times 3}	& \hat{M}_R
 \end{matrix} \right)\simeq  M_0I_6 , 
 \;  V\simeq \dfrac{1}{\sqrt{2}}
 \left(\begin{matrix}
 	-iI_3 	& I_3   \\ 
 	iI_3 	& I_3 
 \end{matrix} \right), \label{eq_UNiss}	
 \end{align}
  meaning that all of six  heavy neutrinos have approximately the same mass $m_{n_i}\simeq M_0$ for all $i>3$.  Hence, the above simple forms of  $M_R$, and $\mu_X$ will result in degenerate heavy neutrino masses, leading to small rates of  lepton flavor violating decays of charged leptons (cLFV) that satisfy the current experimental constraints~\cite{MEG:2016leq, BaBar:2009hkt}.
 Relations in \eqref{eq_RISS} reduce to the following simple forms: 
 \begin{align}
 	\label{eq_mDRISS}
 	 m_D&= M_0\hat{x}_\nu^{1/2}  U^\dagger_{\mathrm{PMNS}},
 	%
 \quad R  =\left( -U_{\mathrm{PMNS}}\frac{\sqrt{\mu_0 \hat{m}_{\nu}}}{M_0},\;  U_{\mathrm{PMNS}}\hat{x}_\nu^{1/2} \right) \simeq \left(0_{3\times 3},\;  U_{\mathrm{PMNS}}\hat{x}_\nu^{1/2} \right),
 \end{align}
where 
\begin{equation}\label{eq_hxnu}
	\hat{x}_\nu\equiv \frac{\hat{m}_\nu}{\mu_0}\equiv \mathrm{diag} \left( \hat{x}_{\nu1}, \hat{x}_{\nu2}, \hat{x}_{\nu3}\right),\; \hat{x}_{\nu a}\equiv\frac{m_{n_a}}{\mu_0},\; a=1,2,3
\end{equation}
 satisfying max$[ |\hat{x}_{\nu a}|]\ll1$ for all $a=1,2,3$.

In our numerical analysis, we will use the best-fit values of the neutrino oscillation data~\cite{ParticleDataGroup:2020ssz} corresponding to the normal order (NO) scheme with $m_{n_1}<m_{n_2}<m_{n_3}$, namely 
\begin{align}
	\label{eq_d2mijNO}
	&s^2_{12}=0.32,\;   s^2_{23}= 0.547,\; s^2_{13}= 0.0216 ,\;  \delta= 218 \;[\mathrm{Deg}] , 
	\crn &\Delta m^2_{21}=7.55\times 10^{-5} [\mathrm{eV}^2], \;
 \Delta m^2_{32}=2.424\times 10^{-3} [\mathrm{eV}^2].
\end{align}

In our numerical analysis, we have used the following relations
\begin{align}
	\label{eq_NO1}
	\hat{x}_{\nu}&= \mu_0^{-1} \mathrm{diag} \left( m_{n_1}, \; \sqrt{  m_{n_1}^2 +\Delta m^2_{21}},\; \sqrt{ m_{n_1}^2 +\Delta m^2_{21} +\Delta m^2_{32}} \right),
	\crn U_{\mathrm{PMNS}} &=\left(
	\begin{array}{ccc}
		c_{12} c_{13} & c_{13} s_{12} & s_{13} e^{-i \delta } \\
		-c_{23} s_{12}-c_{12} s_{13} s_{23} e^{i \delta } & c_{12} c_{23}-s_{12} s_{13} s_{23} e^{i \delta } & c_{13} s_{23} \\
		s_{12} s_{23}-c_{12} c_{23} s_{13} e^{i \delta } & -c_{23} s_{12}  s_{13}e^{i \delta }  -c_{12} s_{23} & c_{13} c_{23} \\
	\end{array}
	\right)  
	\crn&\simeq \left(
	\begin{array}{ccc}
		0.816 & 0.56 & 0.147 e^{-i \delta } \\
		-0.381-0.09 e^{ i \delta } & 0.555 -0.062 e^{ i \delta } & 0.732 \\
		0.418-0.082 e^{ i \delta } & -0.61-0.056 e^{ i \delta } & 0.666 \\
	\end{array}
	\right). 
\end{align}
These numerical values of neutrino masses satisfy the cosmological constraint arising from the Planck 2018 experimental data \cite{Planck:2018vyg}: 
$\sum_{i=a}^{3}m_{n_a}\leq 0.12\; \mathrm{eV}$.  In order to simplify our numerical analysis, we  assume $ m_{n_1} =0.01\; \mathrm{eV} < m_{n_2}<m_{n_3}$.

The other well-known numerical parameters are given in Ref.~\cite{ParticleDataGroup:2020ssz}, namely 
\begin{align}
	\label{eq_ex}
	g &=0.652,\; \alpha_e=\frac{1}{137}= \frac{e^2}{4\pi} ,\; s^2_{W}=0.231,\crn
	m_e&=5\times 10^{-4} \;\mathrm{GeV},\; m_{\mu}=0.105 \;\mathrm{GeV}, \; m_W=80.385 \; \mathrm{GeV}. 
\end{align}
Also the inverted order (IO)  scheme with $m_{n_3}<m_{n_1}<m_{n_2}$ can be considered in the similar way, but the  qualitative results are the same with those from NO scheme, so we will not present here. 

The non-unitary of the active neutrino mixing matrix $\left(	I_3-\frac{1}{2}RR^{\dagger} \right) U_{\mathrm{PMNS}}$ is constrained by other phenomenological aspects such as, for instance, electroweak precision tests,  cLFV decays~\cite{Fernandez-Martinez:2016lgt,  Pinheiro:2021mps, Agostinho:2017wfs}, thus leading to the following constraints 
\begin{align} \label{eq_maxRRd}
\eta\equiv	\frac{1}{2}\left| RR^{\dagger}\right|< \eta_0=
	\begin{pmatrix}
	2\times 10^{-3}	& 3.5\times 10^{-5}  &  8.\times 10^{-3}\\
		3.5\times 10^{-5}&8\times 10^{-4}  & 5.1\times 10^{-3} \\
		8\times 10^{-3}& 5.1\times 10^{-3} & 2.7\times 10^{-3} 
	\end{pmatrix}.
\end{align}
 This constraint is consistent with the data popularly used in recent works discussion on the ISS framework \cite{ Bandyopadhyay:2012px, Dao:2021vqp,Mondal:2021vou}.  The constraint on $\eta$ may be more strict, depending on particular models. For example in the type  III and inverse seesaw models, one has the constraint $|\eta_{aa}|\leq \mathcal{O}(10^{-4})$~\cite{Biggio:2019eeo, Escribano:2021css}. In our numerical analysis, we will choose the values satisfying $|\eta_{33}|\leq 10^{-3}$, which are also consistent with the  updated constraints on neutrino couplings  discussed in Ref.~\cite{ Coutinho:2019aiy,  Manzari:2020eum},   including the constraint from lepton universality previously discussed \cite{deBlas:2013gla}.

 In the next section, we will discuss the model scalar potential as well as  the one-loop contributions to $\Delta a_{\mu,e}$. 
 
 \section{ \label{eq_Deltaamu} Higgs bosons and one-loop contributions to $\Delta a_{e_a}$}
 The Higgs potential satisfying the symmetry mentioned in this work is
\begin{align}
	V&= \mu_1^2 \Phi_1^{\dagger}\Phi_1 + \mu_2^2 \Phi_2^{\dagger}\Phi_2 - \left( \mu_3^2 \Phi^\dagger_2 \Phi_1 + \mathrm{H.c.}\right) 
	\crn & +\dfrac{1}{2} \lambda_1\left(\Phi_1^{\dag}\Phi_1 \right)^2 + \dfrac{1}{2} \lambda_2\left(\Phi_2^{\dag}\Phi_2\right)^2+\lambda_3\left(\Phi_1^{\dag}\Phi_1\right) \left(\Phi_2^{\dag} \Phi_2\right) +\lambda_4\left(\Phi_1^{\dag}\Phi_2\right) \left(\Phi_2^{\dag}\Phi_1\right)  \nonumber\\
	& +\lambda_h\left| {h^+} \right|^4 + \left| {h^+} \right|^2 \left[ \mu_h^2 +\lambda_8 \Phi_1^{\dag}\Phi_1 +\lambda_9\Phi_2^{\dag}\Phi_2  \right] 
	+ \left(\mu\epsilon_{\alpha\beta}\Phi_1^\alpha \Phi_2^\beta h^- + \mathrm{H.c.}\right),\nonumber
\end{align}
where  the $Z_2$ soft breaking term $\left(\mu_3^2\Phi^\dagger_2 \Phi_1 \mathrm{+H.c.}\right)$ is kept in order to generate non-zero mass for the CP-odd neutral Higgs predicted in this model. The quartic term $\left( \Phi^\dagger_2 \Phi_1\right)^2$ vanishes because of the $Z_3$ symmetry, whose assignments for the particle spectrum are given in Table \ref{t_Z23}. Consequently, the one-loop contributions to active neutrino masses discussed  in Ref.~\cite{Ma:2006km}  do not appear in this case,  implying that all entries of $Y^h$  are not constrained by this condition. In general, $\mu_3$, and $\mu$ can be complex, while  the remaining parameters  $\mu_{1,2}$, $\mu_h$, and $\lambda_{1,2,3,4,8,9,h}$ are real \cite{Herrero-Garcia:2017xdu, Davidson:2005cw}. In this work, all of these parameters and vacuum expectation values (vev) are assumed to be real, which corresponds to CP conservation. 

It is easy to find the two minimization conditions of the Higgs potential, then inserting them into the Higgs potential in order to find the physical electrically charged scalars as follows:
\begin{align}
	\begin{pmatrix}
		G^\pm \\H^\pm 
	\end{pmatrix} &=
	\begin{pmatrix}
		\ c_ {\beta} & s_{\beta} \\-s_{\beta}&c_{\beta}
	\end{pmatrix} 
	\begin{pmatrix}
		\phi_1^\pm \\
		\phi_2^\pm
	\end{pmatrix}
	,  \label{eq_phi120} 
	\crn 	\begin{pmatrix}
		h_1^\pm  \\ h_2^\pm
	\end{pmatrix} &=
	\begin{pmatrix}
		s_\varphi & c_\varphi \\ c_\varphi & -s_\varphi
	\end{pmatrix}
	\begin{pmatrix}
		h^\pm \\ H^\pm
	\end{pmatrix}, \quad s_{2\varphi}= \dfrac{\sqrt{2}v\mu}{m^2_{h_2^+}-m^2_{h_1^+}}, 
	\crn 	m^2_{{h_1^+},{h_2^+}} &\equiv \dfrac{1}{2}\left[M^2_{H^+} + M^2_{33} \mp\sqrt{\left(M^2_{H^+} - M^2_{33}\right)^2 + 2v^2\mu^2} \right],
\end{align}
where $t_\beta \equiv \tan\beta=\dfrac{v_2}{v_1}$,    $s_x\equiv \sin x$,  $c_x\equiv \cos x$, $M_{H^+}^2= \frac{\mu_{3}^2}{s_{\beta} c_{\beta}} - \dfrac{1}{2}v^2\lambda_4$,     $M_{33}^2 = \mu_h^2 + \frac{v^2}{2} \left(c_{\beta}^2\lambda_8 + s_{\beta}^2\lambda_9 \right)$.  This result is consistent with the one obtained from the Higgs potential used in Ref~\cite{Herrero-Garcia:2017xdu} after some transformations into a new Higgs basis and parameters.  Namely, the two Higgs doublets $\Phi_i$ are changed into a new basis as follows: 
 \begin{equation}
	\begin{pmatrix}
		H_1 \\H_2 
	\end{pmatrix}=
	\begin{pmatrix}
		\ c_ {\beta} & s_{\beta} \\-s_{\beta}&c_{\beta}
	\end{pmatrix} 
	\begin{pmatrix}
		\Phi_1 \\
		\Phi_2
	\end{pmatrix},  \label{phi12} 
\end{equation} 
 which can be expanded around the minimum as shown below:
 \begin{align}
 	H_1 &=
 	\begin{pmatrix}
 		G^+ \\ \dfrac{v+\varphi^0_1 +iG^0}{\sqrt{2}}
 	\end{pmatrix}  , \; 
 	H_2  =
 	\begin{pmatrix}
 		H^+ \\ \dfrac{\varphi^0_2 +iA}{\sqrt{2}}
 	\end{pmatrix}  \label{btH2}, 
 \end{align}
where $v_1^2+v_2^2=v^2=(246\; \mathrm{GeV})^2$. In addition, the free parameters of the Higgs potential are transformed as follows $ \mu_1^2 c_{\beta}^2 + \mu_2^2 s_{\beta}^2 - \mu_3^2 2s_{\beta}c_{\beta} \to \mu_1^2$,  $\mu_1^2 s_{\beta}^2 + \mu_2^2 c_{\beta}^2 + \mu_3^2 2s_{\beta}c_{\beta} \to \mu_2^2$, $ (\mu_1^2 -\mu_2^2) s_{\beta}c_{\beta} +\mu_3^2 (c^2_{\beta} -s^2_{\beta}) \to \mu_3^2$, $\lambda_8 c^2_{\beta}+\lambda_9 s^2_{\beta} \to \lambda_8 $, $\lambda_8 s^2_{\beta}+\lambda_9 c^2_{\beta} \to \lambda_9 $, $s_{\beta}c_{\beta}(\lambda_9 -\lambda_8)\to \lambda_{10}$, $\mu \to \mu, \;\dots $ The Higgs potential after the transformation \eqref{phi12} takes the form  \cite{Herrero-Garcia:2017xdu}, 
where the physical charged scalar states and their masses are consistent with those ones given in Eq. \eqref{eq_phi120}.

In our numerical calculation, we will use $	m^2_{{h_{k}^+}}$ and the mixing angle  $\varphi$ as free parameters. Three Higgs mass parameters $\mu_2^2$, $\mu_h^2$, and $\mu$ are functions of  the remaining parameters. Thus,
no perturbative limits on the Higgs selfcouplings are necessary to constrain the dependent functions chosen here.  

From the above information we obtain all vertices providing 
one-loop contributions to the $e_b \to e_a \gamma$ decay rates as well as to 
$\Delta a_{e_a}$. They are collected from the lepton Yukawa terms given in 
Eq. \eqref{eq_ylepton1} respecting all symmetries given in Table \eqref{t_Z23}, namely 
\begin{align*}
	\mathcal{L}^\text{yuk}_{\mathrm{lepton}} &=    -\frac{\sqrt{2} m_{e_a}}{v_1} \overline{L_{a}} \Phi^{\dagger}_1 e_{aR}   -Y^N_{k,ba} \overline{N_{bR}}\left(i\sigma_{2} \Phi_k \right)^T L_{a}  - Y^{h}_{(b+3)a}\overline{(N_{(b+3)R})^c} e_{aR} h^+ +\mathrm{h.c.},  
\end{align*}
In order to derive the total neutrino mass matrix in the general ISS form, the leptonic Yukawa terms are rewritten as follows
\begin{align}
	\label{eq_y ffX}
\mathcal{L}^\text{yuk}_{\mathrm{lepton}} &=    -\frac{g m_{e_a}(-t_{\beta})}{ \sqrt{2}m_W} \overline{L_{a}} H_2 e_{aR}   -Y^N_{0,Ia} \overline{N_{IR}}\left(i\sigma_{2} H_2 \right)^T L_{a}  - Y^{h}_{Ia}\overline{(N_{IR})^c} e_{aR} h^+ +\mathrm{h.c.},  
\end{align}
where we denote $Y^N_{0}\equiv  \left( Y^N_{0},\; 0_{3\times 3}\right)^T = -s_{\beta}Y^N_{1} + c_{\beta}Y^N_{2}$ for the ISS mechanism discussed in this work.  It is interesting to link this matrix with $M_D$ given in Eq.~\eqref{eq_MD}, where depending on the $Z_2$ charges of $N_{IR}$  there are two cases where non-zero couplings with only $\Phi_1$ or $\Phi_2$ correspond to $Y^N_{2}=0$ or $Y^N_{1}=0$, respectively.  Namely 
\begin{align} \label{eq_YN0}
	M_D^T= \left( \frac{vf_H^{-1}}{\sqrt{2}}Y^N_{0},\; 0_{3\times 3}\right)^T , \; f_H= \left[\begin{array}{cc}
	 t_{\beta}^{-1}, &  Y^N_{1}=0_{6\times 3}, \; Y^N_{2}\ne 0; \\
		- t_{\beta} ,& Y^N_{2}=0_{6\times 3}, \; Y^N_{1}\ne 0
	\end{array}
	\right. .
\end{align}
 With this new notation, the Yukawa Lagrangian for the THDM type-A is 
\begin{align}
	\label{eq_y ffX1}
	\mathcal{L}^\text{yuk}_{\mathrm{lepton}}= &    \frac{g m_{e_a}t_{\beta}}{ \sqrt{2}m_W} U^{\nu*}_{ai}\overline{n_{iL}}e_{aR}  \left( c_{\varphi} h^+_1 -s_{\varphi} h^+_2 \right)
	%
	+\frac{g f_H}{ \sqrt{2}m_W} U^{\nu}_{(I+3)i} M_{D,ia} \overline{n_{iR}} e_{aL}  \left( c_{\varphi} h^+_1 -s_{\varphi} h^+_2 \right)
	 \crn&- Y^{h}_{Ib}U^{\nu*}_{(I+3)i}\overline{n_{iL}}e_{aR}  \left( s_{\varphi} h^+_1 +c_{\varphi} h^+_2 \right) +\mathrm{h.c.}.  
\end{align}

\begin{figure}[ht]
	\includegraphics[width=12cm]{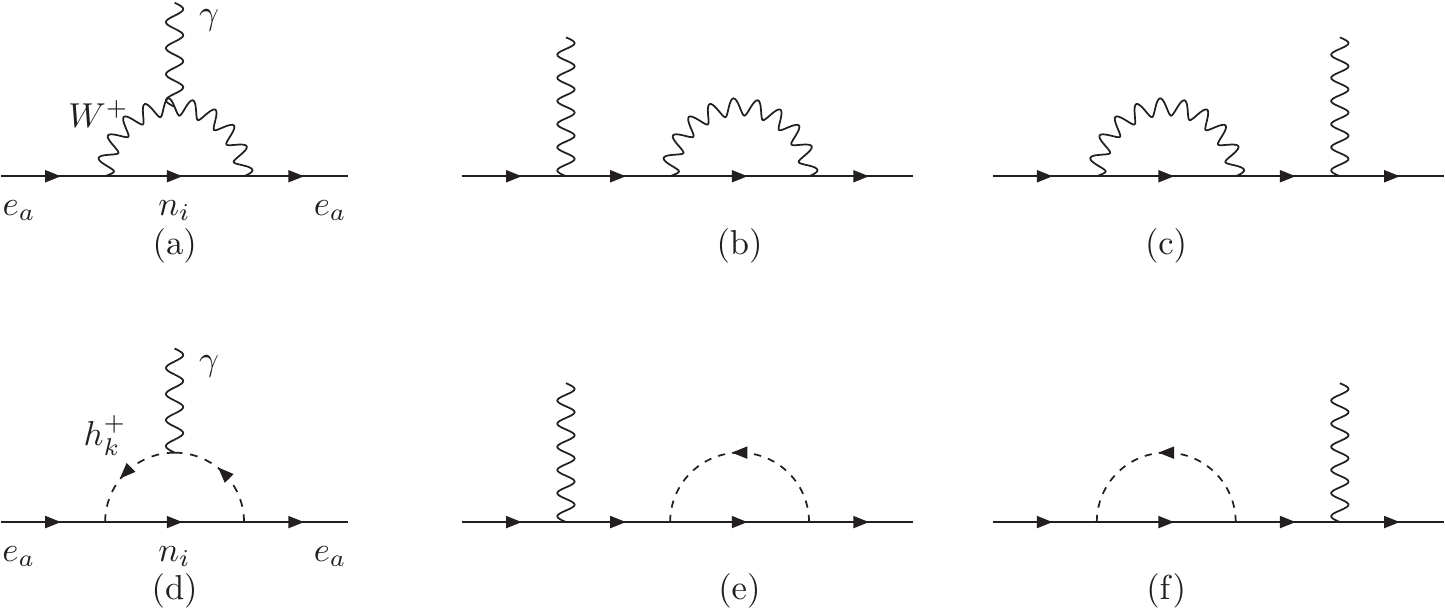}
	\caption{One-loop contributions of $W^\pm$ and $h^\pm_k$ to $(g-2)_{e_a}$  in the unitary gauge, where $k=1,2$. }\label{fig_g2diagram}
\end{figure}

Then, all relevant couplings  are given in the following lepton Yukawa interaction   Lagrangian
\begin{align}
	\label{eq_g2Lagrangian}
	\mathcal{L}= \sum_{a=1}^3 \sum_{i=1}^{9}\left[  \frac{g}{\sqrt{2} m_W} \sum_{k=1}^2\overline{n_i} \left(  	\lambda^{L,k}_{ia}  P_L +\lambda^{R,k}_{ia}P_R \right) e_a h^+_k   +\frac{g}{\sqrt{2}} U^{\nu*}_{ai} \overline{n_{i}}\gamma^{\mu}P_L e_aW^+_{\mu}  \right]
	%
	 +\mathrm{h.c.},
\end{align}
where
\begin{align}
	\label{eq_lakLR}
	\lambda^{L,1}_{ia}&= \sum_{I=1}^6 f_HM_{D,Ia}c_{\varphi}U^{\nu}_{(I+3)i} \simeq f_H c_{\varphi}\times \left[\begin{array}{cc}
	0,	& \quad i\leq 3 \\
		\left( M^T_{D}V\right) _{a(i-3)},	& \quad i > 3 
	\end{array}\right., 
\crn \lambda^{L,2}_{ia}& \simeq	-\lambda^{L,1}_{ia}t_{\varphi}= \lambda^{L,1}_{ia} \left[  c_{\varphi}\to -s_{\varphi}\right],
	\crn	\lambda^{R,1}_{ia}&= m_{e_a}t_{\beta} c_{\varphi} U^{\nu*}_{ai} - \sum_{I=1}^6 \frac{v}{\sqrt{2}}  s_{\varphi} Y^{h}_{Ia}U^{\nu*}_{(I+3)i} 
	\crn &\simeq  \left[\begin{array}{cc}
		m_{e_a}t_{\beta} c_{\varphi} \left[ U^*_{\mathrm{PMNS}}\left( I_3 -\frac{1}{2}\hat{x}_{\nu} \right) \right]_{ai} +\frac{vs_{\varphi}}{\sqrt{2}}\left( Y^{hT}R^{T}U^*_{\mathrm{PMNS}}\right) _{ai},	& \quad i\leq 3 \\
		m_{e_a}t_{\beta} c_{\varphi} \left( RV\right)^*_{a(i-3)}-	\frac{vs_{\varphi}}{\sqrt{2}}\left[Y^{hT} \left(I_6- \frac{1}{2}R^{T}R^*\right)V^*\right]_{a(i-3)}	& \quad i > 3 
	\end{array}\right.,
	\crn	\lambda^{R,2}_{ia}&= -m_{e_a}t_{\beta} s_{\varphi} U^{\nu*}_{ai} - \sum_{I=1}^6 \frac{v}{\sqrt{2}}  c_{\varphi} Y^{h}_{Ia}U^{\nu*}_{(I+3)i}
	\crn &\simeq  \left[\begin{array}{cc}
		-m_{e_a}t_{\beta} s_{\varphi} \left[  U^*_{\mathrm{PMNS}} \left( I_3 -\frac{1}{2}\hat{x}_{\nu}\right)  \right]_{ai} +\frac{v c_{\varphi}}{\sqrt{2}}\left( Y^{hT}R^{T}U^*_{\mathrm{PMNS}}\right) _{ai},	& \quad i\leq 3 \\
		-m_{e_a}t_{\beta} s_{\varphi} \left( RV\right)^*_{a(i-3)}-	\frac{vc_{\varphi}}{\sqrt{2}}\left[Y^{hT} \left(I_6- \frac{1}{2}R^{T}R^*\right)V^*\right]_{a(i-3)}	& \quad i > 3 
	\end{array}\right. 
\crn &= \lambda^{R,1}_{ia} \left[ s_{\varphi}\to c_{\varphi},\; c_{\varphi}\to -s_{\varphi}\right]. 
\end{align}
The Feynman diagrams giving one-loop contributions to $(g-2)_{e_a}$ corresponding to the  Lagrangian \eqref{eq_g2Lagrangian} are shown in Fig. \ref{fig_g2diagram}. 
 
We do not list here the couplings of neutral gauge and Higgs bosons because they give suppressed contributions to  $a^{\mathrm{NP}}_{e_a}$. In particular, the relevant couplings are only the ones with usual charged leptons $s^0\overline{e_a}e_a$ and $Z_{\mu}\overline{e_a}\gamma^{\mu}e_a$. The one-loop level contribution arising from the $Z$ gauge boson exchange is the same as the predicted by the SM. The contributions arising from new neutral Higgs bosons are not larger than the one coming from the SM-like Higgs boson since they are suppressed by a factor of the order of $\mathcal{O}(10^{-14})$, because we assume here their masses are at the TeV scale. 

We will use the approximation that  $ m^2_{n_i}/m_W^2=0$ with $i\le 3$ and $m^2_{n_i}/m_W^2=M_0^2/m_W^2=x_W$ with $i>3$. Then, 
the contribution to $a_{e_a}$ arising from the $W$ exchange has the form: 
\begin{align}
	\label{eq_Deltaaea}
	 a_{e_a}(W)&= -\frac{g^2 m^2_{e_a}}{8\pi^2 m^2_W}  \left[  -\frac{5}{12} +  \left( R^*R^T\right)_{aa} \times \left(\tilde{f}_V \left(x_W\right)+ \frac{5}{12}\right) \right],
\end{align}
where 
\begin{align}
	\label{eq_fVx}
	\tilde{f}_V(x)&=	\frac{-4x^4 +49x^3 -78 x^2 +43x -10 -18x^3\ln x}{24(x-1)^4}, 
	\crn & \tilde{f}_V(0)= -\frac{5}{12}\leq \tilde{f}_V(x)\leq 	\tilde{f}_V(\infty)= -\frac{1}{6}.
\end{align}
Because $|\tilde{f}_V \left(x_W\right)+ \frac{5}{12}|\leq \frac{5}{12}$ and $\left( R^*R^T\right)_{aa} \leq \mathcal{O}(10^{-3})\ll1$, Eq.~\eqref{eq_Deltaaea}  equals to the one-loop level contribution predicted by the SM, see example in Ref.~\cite{Jegerlehner:2009ry}: 
\begin{equation}\label{eq_amuSMW}
a^{(1)\mathrm{SM}}_{\mu}(W) =\frac{g^2 m^2_{\mu}}{8\pi^2 m^2_W}\times \frac{5}{12}\simeq 383\times 10^{-11},\; 	 \frac{g^2 m^2_{\mu}}{8\pi^2 m^2_W}\simeq 9.19\times 10^{-9}.
\end{equation}

 The one-loop level contribution to $\Delta a_{e_a}$ arising from the exchange of the electrically charged scalar singlet $h^\pm_k$ is given by  \cite{Crivellin:2018qmi}:
\begin{align}
	\label{eq_aHpm}
	a_{e_a} (h^\pm_k)& =- \frac{g^2m_{e_a}\;}{8 \pi^2 m^2_W}     \sum_{i=1}^{9} \frac{\lambda^{L,k*}_{ia } \lambda^{R,k}_{ia }m_{n_i} f_{\Phi}(x_{i,k}) + m_{e_a} \left(  \lambda^{L,k*}_{ia } \lambda^{L,k}_{ia } + \lambda^{R,k*}_{ia } \lambda^{R,k}_{ia }\right)  \tilde{f}_{\Phi}(x_{i,k})}{m^2_{h^\pm_k}} , 
\end{align}
where $x_{i,k}\equiv m^2_{n_i}/m^2_{h^\pm_k}$ and  the loop functions appearing in Eq. (\ref{eq_aHpm}) have the forms:
\begin{align}
	\label{eq_MasterFunc}
	f_\Phi (x)&= \frac{x^2-1 -2x\ln x}{4(x-1)^3}, \quad 
%
\tilde{f}_\Phi(x)  = \frac{2x^3 +3x^2 -6x +1 -6x^2 \ln x}{24(x-1)^4}, 
\crn & f_{\Phi}(\infty)= 0\leq f _{\Phi}(x)\leq 	f_{\Phi}(0)=  \frac{1}{4},
%
\;   \tilde{f}_{\Phi}(\infty)= 0 \leq \tilde{f}_{\Phi}(x)\leq 	\tilde{f}_{\Phi}0)= \frac{1}{24}. 
\end{align}  
And the deviation from the SM is defined as follows:
\begin{align}
	\Delta a_{e_a}= \sum_{k=1}^2  a_{e_a}(h^\pm_k) +  \Delta a_{e_a}(W), \;  \Delta a_{e_a}(W)\equiv a_{e_a}(W)-a^{(1)\mathrm{SM}}_{e_a}(W), 
\end{align}
where   $a^{(1)\mathrm{SM}}_{\mu}(W) \simeq 3.83 \times 10^{-9}$~\cite{Jegerlehner:2009ry}.

Using the approximations $m^2_{n_i}/m^2_{h^\pm_k}\simeq0$ for $i\leq3$ and  $m^2_{n_i}/m^2_{h^\pm_k}\simeq M_0^2/m^2_{h^\pm_k}=x_k$, we have $ f_{\Phi}(x_{i,k})\simeq f_{\Phi}(0)$ for $i\leq 3$ and $ f_{\Phi}(x_{i,k})\simeq f_{\Phi}(x_k)$ for $i>3$.  Following Eqs. \eqref{eq_UNiss} and  \eqref{eq_mDRISS}, we obtain that the one loop level contribution to $a_{e_a}$ due to the exchange of  $h^\pm_1$ is given by
\begin{align}
	a_{e_a}(h^\pm_1)& = - 9.19\times 10^{-9} \times \frac{m^2_{e_a}}{m^2_{\mu}}
\crn&	\times \mathrm{Re}\left\{ \sum_{c=1}^{3}  f_H  \left[ c^2_{\varphi} t_{\beta}\left|U_{\mathrm{PMNS},ac} \right|^2 \frac{m_{n_c}}{\mu_0} - \frac{v c_{\varphi}s_{\varphi}}{\sqrt{2}m_{e_a}}U_{\mathrm{PMNS},ac} \left(\frac{m_{n_c}}{\mu_0} \right)^{1/2} Y^{h}_{(c+3)a} \right]x_1f_{\Phi}(x_1)
	\right.\crn&  \quad +  \sum_{c=1}^{3}  \left[ \left|U_{\mathrm{PMNS},ac} \right|^2 \left(f_H^2 c^2_{\varphi} \frac{m_{n_c}}{\mu_0} \right) \right] x_1 \tilde{f}_{\Phi}(x_1)
	\crn & \quad+ \frac{m^2_{e_a} t_{\beta}^2c_{\varphi}^2}{m^2_{H^\pm_1}} \left[ \frac{1}{24} - \sum_{c=1}^{3} \left[ \left|U_{\mathrm{PMNS},ac} \right|^2\frac{m_{n_c}}{\mu_0} \right] \left( \frac{1}{24} -\tilde{f}_{\Phi}(x_1) \right) \right]
	\crn & \quad+ \frac{v^2s_{\varphi}^2}{2 m^2_{H^\pm_1}} \left[ \sum_{c=1}^{3} \left[ \left|Y^{h}_{(c+3)a} \right|^2 \frac{m_{n_c}}{\mu_0} \right]  \left(  \frac{1}{24} -\tilde{f}_{\Phi}(x_1)  \right)  + \left(Y^{h\dagger}Y^{h}\right)_{aa} \tilde{f}_{\Phi}(x_1)\right]
	\crn& \quad-\left.  \frac{vm_{e_a}t_{\beta} s_{2\varphi}}{\sqrt{2}  m^2_{H^\pm_1}} \left( \frac{1}{24}- \tilde{f}_{\Phi}(x_1)  \right) \sum_{c=1}^{3} \left[ U_{\mathrm{PMNS},ac}Y^{h}_{(c+3)a} \left(\frac{m_{n_c}}{\mu_0} \right)^{1/2}\right]  	\right\},	\label{eq_Hpm1}
	\\  a_{e_a}(h^\pm_2)&= a_{e_a}(h^\pm_1) \left[ x_1 \to\; x_2,\;s_{\varphi}\to  c_{\varphi}, \; c_{\varphi} \to -s_{\varphi}\right].	\label{eq_Hpm2}
\end{align}
In the real part of Eq. \eqref{eq_Hpm1}, the first line corresponds to 
the chirally-enhanced part proportional to $\lambda^{L,1*}\lambda^{R,1}$ whereas the second and remaining lines are the parts proportional to $\lambda^{L,1*}\lambda^{L,1}$ and 
$\lambda^{R,1*}\lambda^{R,1}$, respectively.

Now we compare our results given in Eq.~\eqref{eq_Hpm1} with the one-loop contribution due to the exchange of  singly electrically charged Higgs bosons in the original versions~\cite{Aoki:2009ha} without $N_{IR}$ and the singlet $h^{\pm}$. Now we assume $h^{\pm}_1\equiv H^\pm $ in Eq.~\eqref{eq_phi120}, corresponding to $c_{\varphi}=1, s_{\varphi}=0$. The absence of $N_{IR}$ can be conveniently derived from  $f_H=0$ and $f(x_1)=0$, thus implying that
 the only one-loop contribution from $h^{\pm}_1$ only consists of the first term in the third line of the real part given in Eq. \eqref{eq_Hpm1}, which is proportional to $-9.19\times 10^{-9}\times  m^4_{e_a} t_{\beta}^2c_{\varphi}^2/(24 m^2_{\mu}m^2_{H^\pm_1})$, which yields a small and negative contribution to $\Delta a^{\mathrm{NP}}_{\mu}$ \cite{Jueid:2021avn}. The dominant contributions to AMM arise  from two-loop Barr-Zee type diagrams. 
 The similar conclusion for the THDM type-B where $t^2_\beta\to t^{-2}_\beta$, hence has suppressed two-loop contributions to AMM for $t_{\beta}\ge0.4$. 

If the mixing between two singly charged Higgs bosons vanishes, namely $s_{\alpha}c_{\alpha}=0$, then all of the remaining terms in both Eqs.~\eqref{eq_Hpm1} and \eqref{eq_Hpm2} are negative, thus not allowing to accommodate the experimental data on muon and electron anomalous magnetic moments.

Using the constraint \eqref{eq_maxRRd} for $RR^{\dagger}=U_{\mathrm{PMNS}} \hat{x}_{\nu}U_{\mathrm{PMNS}}^{\dagger}$ we have $\hat{x}_{\nu}<\mathcal{O}(10^{-3})$. Therefore, we will choose a safe upper bound  as follows 
\begin{equation}
	\label{eq_maxhxnu}
 \hat{x}_{\nu3}\equiv \left(\hat{x}_{\nu}\right)_{33}=\frac{m_{n_3}}{\mu_0}\leq 10^{-3}. 
\end{equation}

To avoid unnecessary independent parameters of $Y^{h}_{Ia}$ without any qualitative  AMM results discussed on this work and in order to cancel large one-loop contributions from these Higgs bosons to the cLFV decays $e_b\to e_a\gamma$, we assume that
\begin{align}
	\label{eq_Yh31}
	\sum_{c=1}^3U_{\mathrm{PMNS},ac} \left(\frac{m_{n_c}}{m_{n_3}} \right)^{1/2} Y^{h}_{(c+3)b}=Y^{d}_{a}\delta_{ab}\,.
\end{align}
The total one-loop level contribution arising from the exchange of two singly charged Higgs bosons is written as
\begin{align}
a_{e_a}(h^\pm)&\equiv 	a_{e_a}(h^\pm_1)+a_{e_a}(h^\pm_2) 
%
=a_{e_a,0}(h^\pm)+\dots, \label{eq_Hpm}
 \\ a_{e_a,0}(h^\pm)& = 9.19\times 10^{-9}  f_H  \times\frac{m_{e_a}^2}{m^2_{\mu}}
%
\times  \mathrm{Re}\left\{\frac{vs_{2\varphi}}{2\sqrt{2}m_{e_a}} \hat{x}_{\nu3}^{1/2} Y^{d}_{a} \left[ x_1f_{\Phi}(x_1) -x_2f_{\Phi}(x_2)\right]
\right\}, \label{eq_Hpm0}
\end{align}
where $a_{e_a,0}(h^\pm)$ denotes the dominant term of the chirally-enhanced part coming from the second one in the first line of the real part given in Eq.~\eqref{eq_Hpm1} where $x_k$ is the part relating with the contribution from $h^{\pm}_k$ exchange. This conclusion can be qualitatively understood from  the property of the large factor $v/(2\sqrt{2}m_{e_a})$ as well as large free Yukawa couplings up to the perturbative limit max[$|Y^{h}_{Ia}|]\le \sqrt{4\pi}$. In addition, the sign of this term can be the same as $\Delta a^{\mathrm{NP}}_{e_a}$ depending on the sign of Re[$Y^{d}_{a}$] when all other factors are fixed. As a result, this term  can easily explain both signs of $\Delta a^{\mathrm{NP}}_{\mu}$ and $\Delta a^{\mathrm{NP}}_{e}$ that are still in conflict between different experimental results and needed to be confirmed in the future. Our numerical investigation showed that the term in Eq.~\eqref{eq_Hpm0} is the dominant one, and the sum of all the remaining terms is suppressed in the allowed regions of parameter space. More detailed estimations confirming that the  remaining contributions are suppressed were given in Ref.~\cite{Hue:2021zyw}. Finally, $a_{e_a,0}(h^\pm)\neq 0$ only when the mixing between two $SU(2)_L$  doublets and singlets is non-zero $s_{\varphi}c_{\varphi}\ne0$, and their masses are non degenerate $m_{h^\pm_1}\ne m_{h^\pm_2}$. 

We comment here an important property that  $a_{e_a,0}(h^\pm)$ in Eq. \eqref{eq_Hpm0} keeps the same form for both types of THDM  A and B, because these models control  only the first term of  $\lambda^R_{ia}$ in Eq. \eqref{eq_lakLR}, which is proportional to $t_{\beta}$ or $t^{-1}_{\beta}$. This key factor  controls  loop contributions to AMM, which are proportional to suppressed power of $t_{\beta}^{-1}$ with large $t_{\beta}$ in  THDM of type-B, including the model type-I. Hence, it is impossible to accommodate the AMM data in the original version of the THDM type I. On the other hand, the THDM type-A has one and two loop contributions to AMM consisting of much enhanced factors of $t^2_{\beta}$ and $t^4_{\beta}$, respectively. Therefore, original versions  can predict large loop contributions to AMM, provided that the new Higgs bosons are light having low masses of about few hundred GeV.  These might be excluded by future collider experiments.
Then, the presence of  $a_{e_a,0}(h^\pm)$ is an alternative way to explain the AMM data. 

\begin{figure}[ht]
	\centering\begin{tabular}{cc}
		\includegraphics[width=7.5cm]{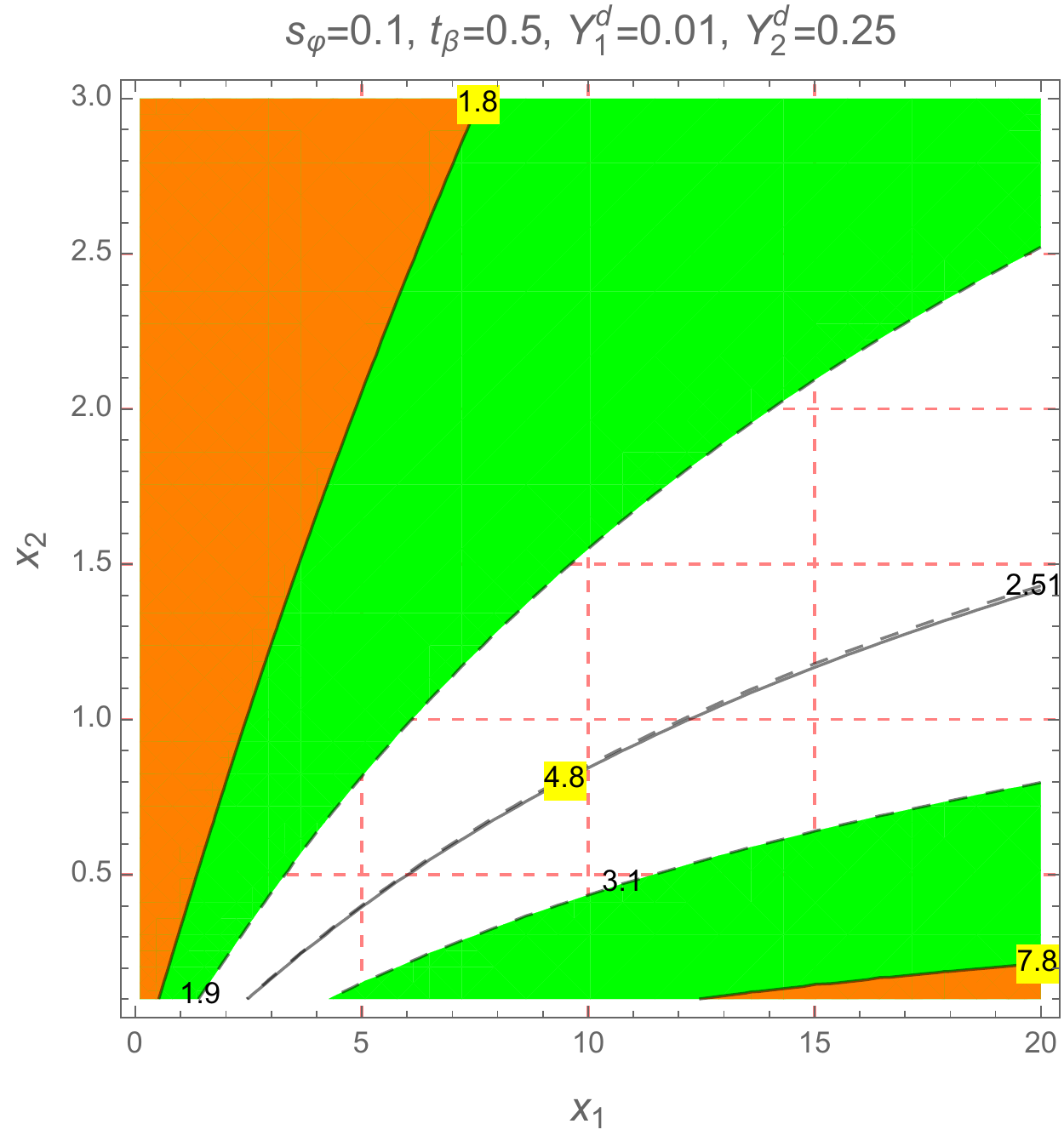}&	\includegraphics[width=7.5cm]{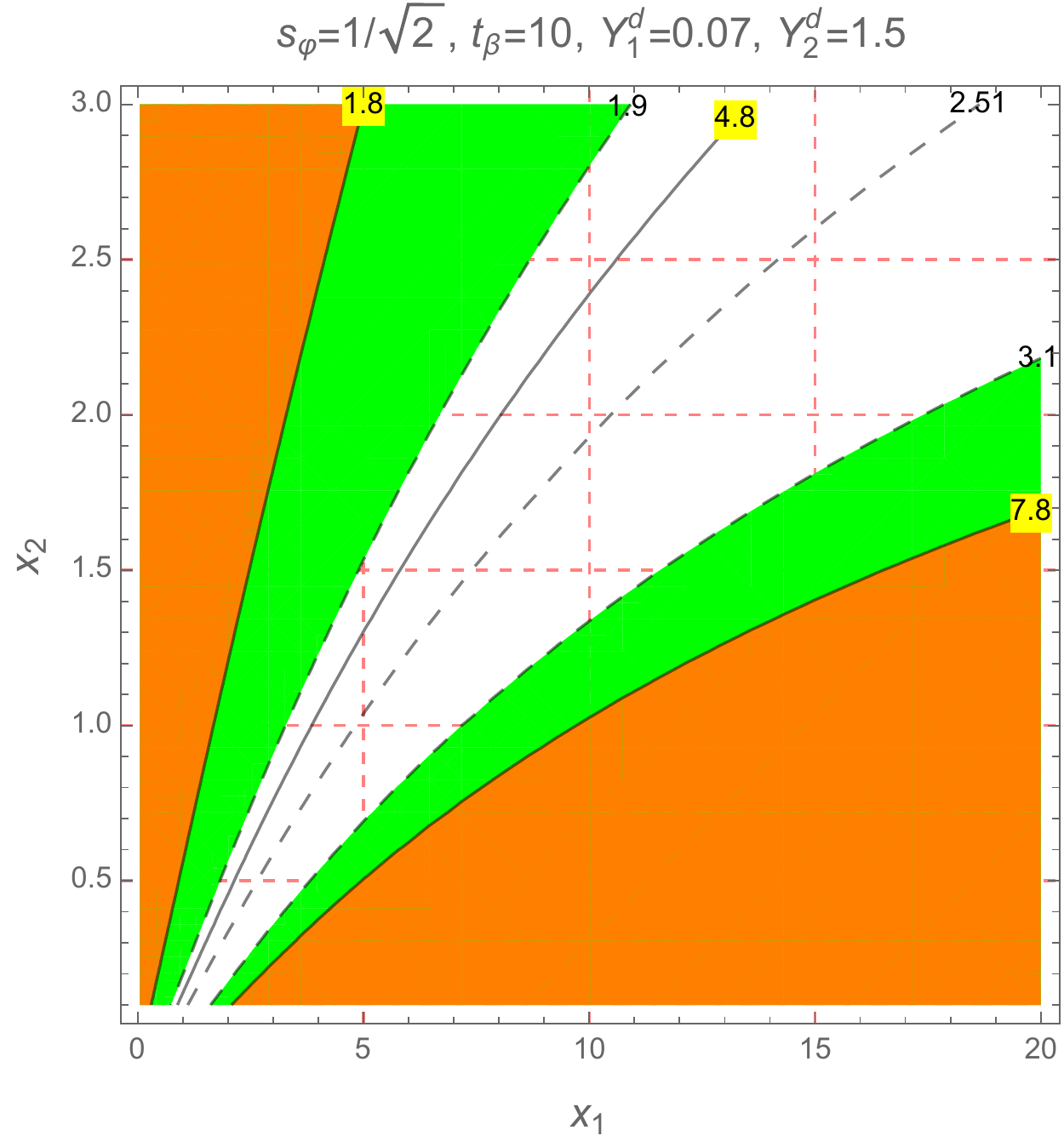}\\	
	\end{tabular}
	\caption{ Contour plots of $\Delta a_{\mu}(h^{\pm})\times 10^{9}$ and $ \Delta a_{e}(h^{\pm})\times 10^{13} $ as functions of $x_1$ and $x_2$, where $\hat{x}_{\nu3}=10^{-3}$ and $f_H=t_{\beta}^{-1}$. The green (orange) regions are excluded by the  1$\sigma$ data of $\Delta a^{\mathrm{NP}}_{\mu}$ ($\Delta a^{\mathrm{NP}}_{e}$). The black and dashed-black curves show the constant values of  $\Delta a_{e}(h^{\pm})\times 10^{13}$  and $\Delta a_{\mu}(h^{\pm})\times 10^{9}$, respectively. }\label{fig_muae1}
\end{figure}

Now we consider the case of $f_H=t_{\beta}^{-1}$, which corresponds to  
the models where the RH neutrino singlets couple with $\Phi_2$, whereas the charged leptons couple with $\Phi_1$ \cite{Li:2020dbg, Hue:2021zyw}. Now, increasing values of $a_{e_a,0}(h^\pm)$ in Eq. \eqref{eq_Hpm0} require small $t_{\beta}$ values, thus implying that 
the scanning range of $t_{\beta}$ should be chosen from the lower bound $t_{\beta}\ge 0.4$. Numerical illustrations of $a_{e_a,0}(h^\pm)$ are shown in Fig. \ref{fig_muae1} with fixed $\hat{x}_{\nu3}=10^{-3}$, i.e.,  $0\neq |(Y^N_2)_{ab}|<\sqrt{4\pi}$. 
 In addition, there are different fixed values of $s_{\varphi}, t_{\beta},Y^{d}_{1}$, and $Y^{d}_{2}$ shown in the respective panels, namely   small $t_{\beta}=0. 5$ and large $t_{\beta}=10$. We have checked that $a_{e_a,0}(h^\pm) \simeq a_{e_a}(h^\pm)$. Now we  estimate the allowed ranges of $M_0$, $m_{h^\pm_{{1,2}}}$, which are affected from the perturbative limit of the Yukawa coupling matrix  $Y^N_0$ defined in Eq.~\eqref{eq_YN0} and related with $M_0$ through Eqs.  \eqref{repara}, \eqref{eq_MD},  and \eqref{eq_mDRISS}. We have two different  constraints corresponding to $f_H=t^{-1}_{\beta}$ and $(-t_{\beta})$.  The constraint for  $f_H=t^{-1}_{\beta}$ is 
 \begin{align} \label{eq_YN0bound}
 	M_0 \left| \left( \hat{x}_{\nu}^{1/2}U^{\dagger}_{\mathrm{PMNS}}\right)_{ab}\right|= \frac{v t_{\beta}}{\sqrt{2}} \left| c_{\beta} (Y^N_2)_{(a+3)b}\right|< \frac{v s_{\beta}}{\sqrt{2}}\sqrt{4\pi}. 
 \end{align}
We can choose a more strict upper bound of $M_0 \leq \hat{x}_{\nu3}^{-1/2}s_{\beta}v\sqrt{\pi/2}= 9.7 s_{\beta}$ TeV for $f_H=t_{\beta}^{-1}$ and $ \hat{x}_{\nu3}=10^{-3}$.  Therefore for $ \hat{x}_{\nu3}\leq 10^{-3}$ and  $t_{\beta}\geq 0.4$, values of $M_0\leq 6.4$ TeV are always acceptable. Applying this constraint, we consider a benchmark where 
$M_0=3$ TeV in order to estimate the  allowed values of $m_{h^\pm_{1,2}}$. We can see that in the left-panel of figure \ref{fig_muae1},  the range $0.2\leq x_2=M_0^2/m^2_{h^\pm_2}\leq 1.5$ with $x_1=M_0^2/m^2_{h^\pm_1}=10$ is allowed with respect to $2.5\; \mathrm{TeV}\leq m_{h^{\pm}_2}\leq  5$ TeV and $m_{h^{\pm}_1}\simeq 0.95$ TeV. Similarly for the right panel of figure \ref{fig_muae1}, we can choose $M_0=$ 5 TeV and larger $ m_{h^\pm_2}=M_0/\sqrt{2.5}$ and  $m_{h^{\pm}_1}\geq 5/\sqrt{15}>1$ TeV.  

\begin{figure}[ht]
	\centering\begin{tabular}{cc}
		\includegraphics[width=8.cm]{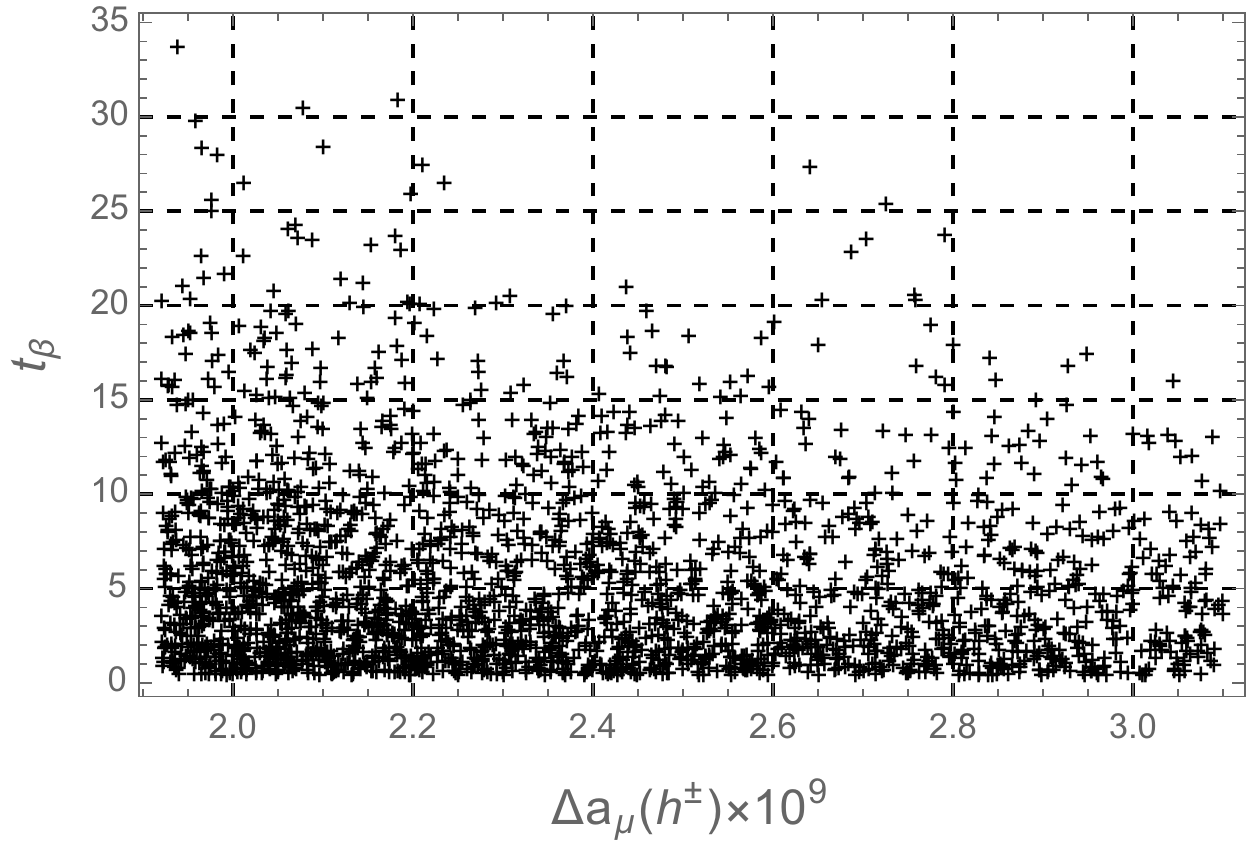}
	&	\includegraphics[width=8.cm]{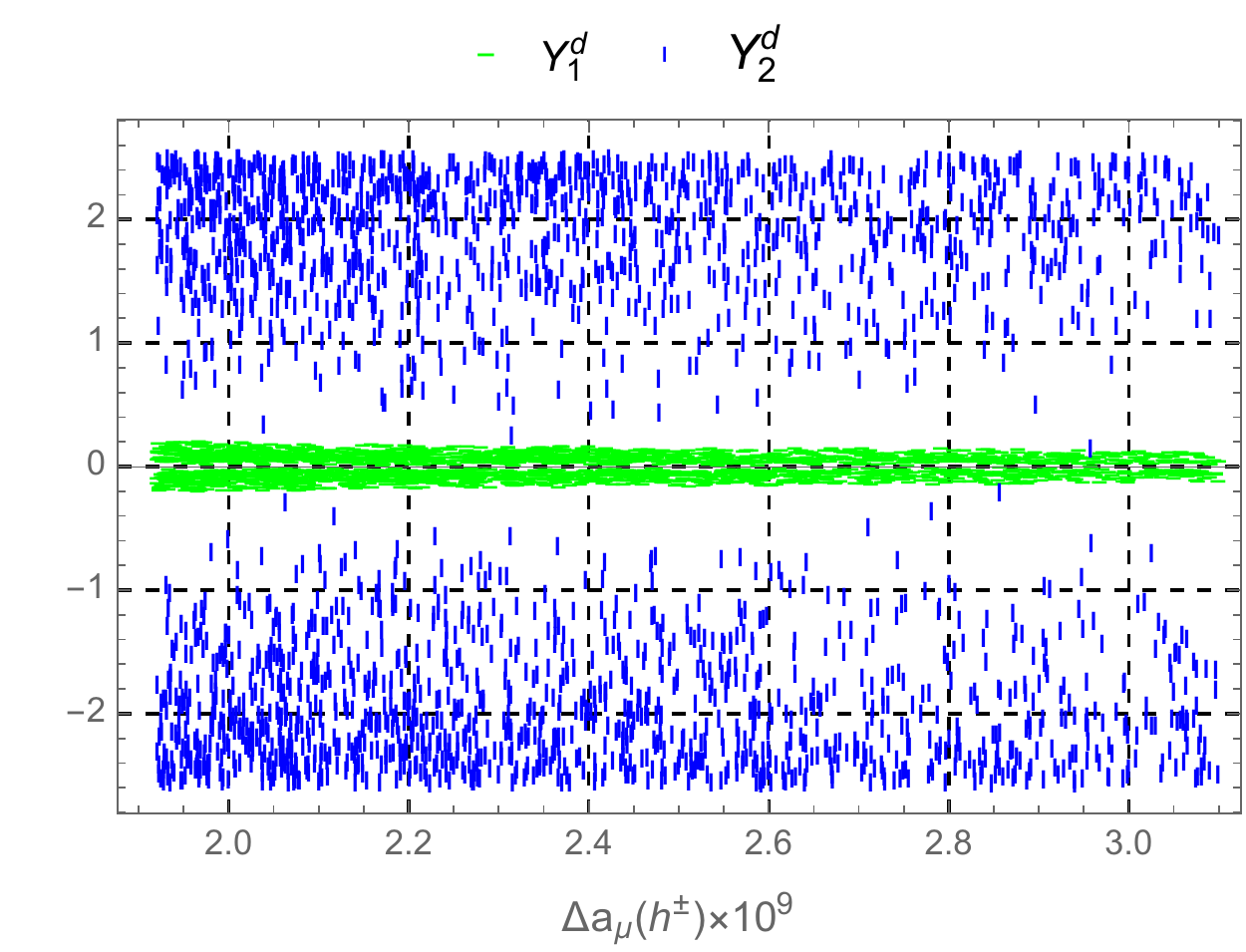}\\
	\end{tabular}
	\caption{ The correlations of $\Delta a_{\mu}(h^\pm)$ vs $ t_{\beta}$ and $Y^d_{1,2}$ with $f_H=t_{\beta}^{-1}$.  }\label{fig_amuvsX}
\end{figure}

In the last discussion we will focus on the allowed regions consisting  of masses of  heavy neutrinos and singly charged Higgs bosons below few TeV so that they can be detected by future colliders. The allowed regions are  defined as they result in the two values of  $ \Delta a_{\mu}(h^\pm)$ and $ \Delta a_{e}(h^\pm)$ both satisfying the experimental data of the muon and electron anomalous magnetic moments within the 1 $\sigma$ level, and all perturbative limits of the Yukawa couplings $Y^{h}_{Ia}$ and $Y^N_{Ia}$ are satisfied, namely $|Y^h_{Ia}|,|Y^N_{2,Ia}|<\sqrt{4\pi}$. The region  of parameter space used to scan  is chosen as follows:
\begin{align}
	\label{eq_scanRanges}
	&  0.8\;  \mathrm{TeV} \leq m_{h^\pm_{1}}, m_{h^\pm_{2}}\le5 \; \mathrm{TeV};   10 \; \mathrm{GeV}\le  M_0\le \; 5\;  \mathrm{TeV}; 
	\crn &0.3 \le t_{\beta}\leq50;\;  \left| s_{\varphi}\right|\le 1; \; |Y^{d}_{ab}|\le3;\; 10^{-7}\leq \; \hat{x}_{\nu3}\equiv \left( \hat{x}_{\nu}\right)_{33}=\frac{m_{n_3}}{\mu_X} \leq 10^{-3}.
\end{align}
Here we fix $m_{n_1}=0.01$ eV corresponding to the NO scheme used in our numerical analysis. 
Smaller values of $m_{n_1}$ will result in small allowed ranges of $Y^d_{1,2}$ because of the perturbative limit affecting the relation given in Eq. \eqref{eq_Yh31}. The scanned range of  $\hat{x}_{\nu3}$ satisfies the non-unitary constraint given in Eq.~\eqref{eq_maxRRd}. The numerical results confirm that $\left|	a_{e_a,0}(h^\pm)/	a_{e_a}(h^\pm) \right|\simeq 1$, namely $0.995 <\left|a_{e,0}(h^\pm)/	a_{e}(h^\pm) \right|< 1.005$ and $1\leq \left|a_{\mu,0}(h^\pm)/	a_{\mu}(h^\pm) \right|\leq 1.03$. Hence the discussion about correlations between different contributions in Eq.~\eqref{eq_Hpm2} will not be shown. The correlations between important free parameters vs.  $\Delta a_{\mu}(h^\pm)$ are shown in Fig. \ref{fig_amuvsX}. 
 As mentioned above, large $t_{\beta}$ favors small $\Delta a_{\mu}(h^\pm)$, leading to an upper bound on $t_{\beta}$ in the allowed regions, see the left-panel of Fig.~\ref{fig_amuvsX}.  
The scanned ranges in Eq. \eqref{eq_scanRanges} allow all experimental ranges of $\Delta a_{e,\mu}$.  In addition,  the dependence between $\Delta a_{e}$ and $\Delta a_{\mu}$ is not interesting. We know instead that $\Delta a_{e}(h^\pm)\simeq \Delta a_{e,0}(h^\pm)$ is a function of  the  Yukawa coupling $Y^d_{1}$.  Hence, the dependence of  $\Delta a_{e}(h^\pm)$ on $\Delta a_{\mu}(h^\pm)$ can be seen from the dependence of the  Yukawa coupling $Y^d_{1}$  in the right panel,  where it is bounded in a more restrictive range than the one given in Eq.~\eqref{eq_scanRanges}, see Table \ref{t_alloewitb}, where other allowed ranges are also listed. 
\begin{table}[ht]
	\centering 
	\begin{tabular}{c|cccccc}
		
		&$t_{\beta }$ & $|s_{\varphi }|$ & $M_0$ [GeV] & $ \left|Y_{1}^d\right|$ & $ \left|Y_{2}^d\right|$ &$ \hat{x}_{\nu3}$\\
		\hline 
		Min& 0.4& 0.03 & 282 &   0.003 & 0.171& $2.92\times 10^{-7}$ \\ 
		Max &33.7 & 0.999 & 5000  & 0.203 & 2.529 & $10^{-3}$ \\ 
	\end{tabular}
	\caption{Allowed ranges corresponding to the scanned region given in Eq.~\eqref{eq_scanRanges}, considered in case of $f_H=t_{\beta}^{-1}$. }\label{t_alloewitb}
\end{table}

\begin{figure}[ht]
	\centering\begin{tabular}{cc}
		\includegraphics[width=8.cm]{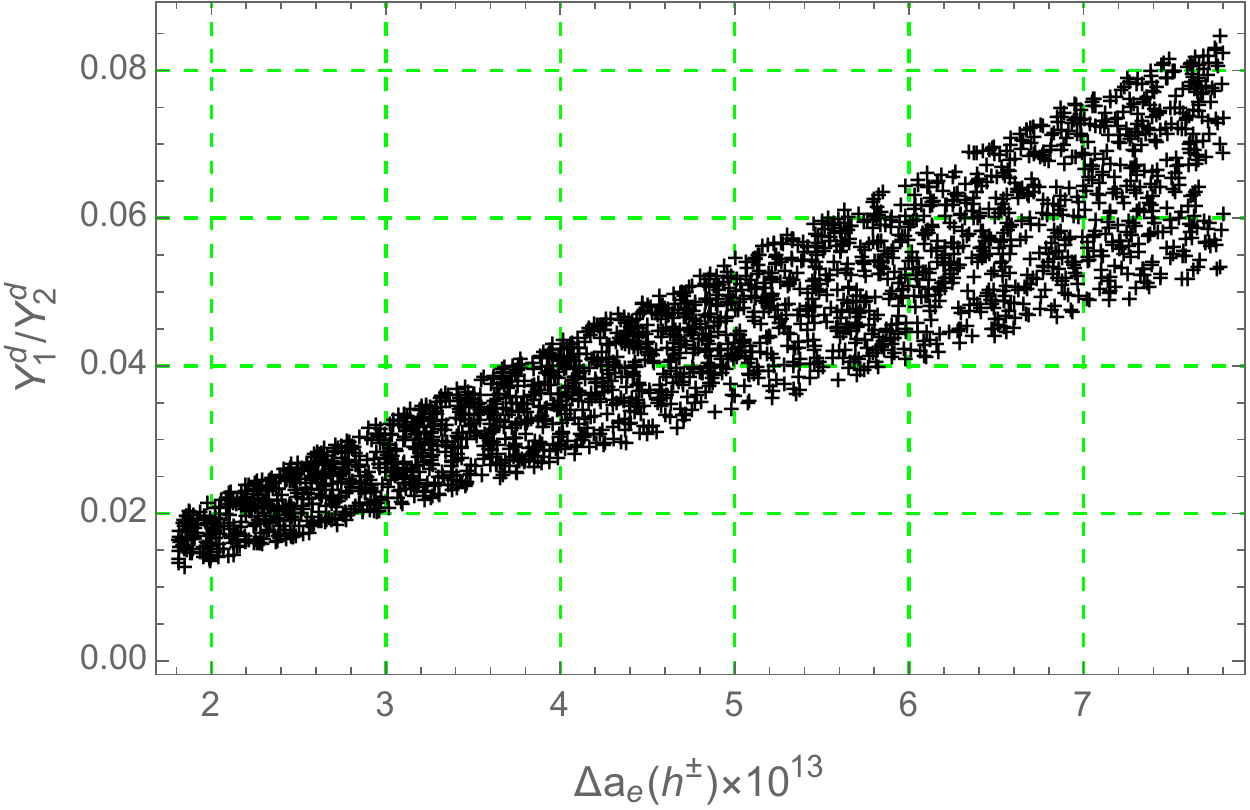}
		&	
		\includegraphics[width=8.cm]{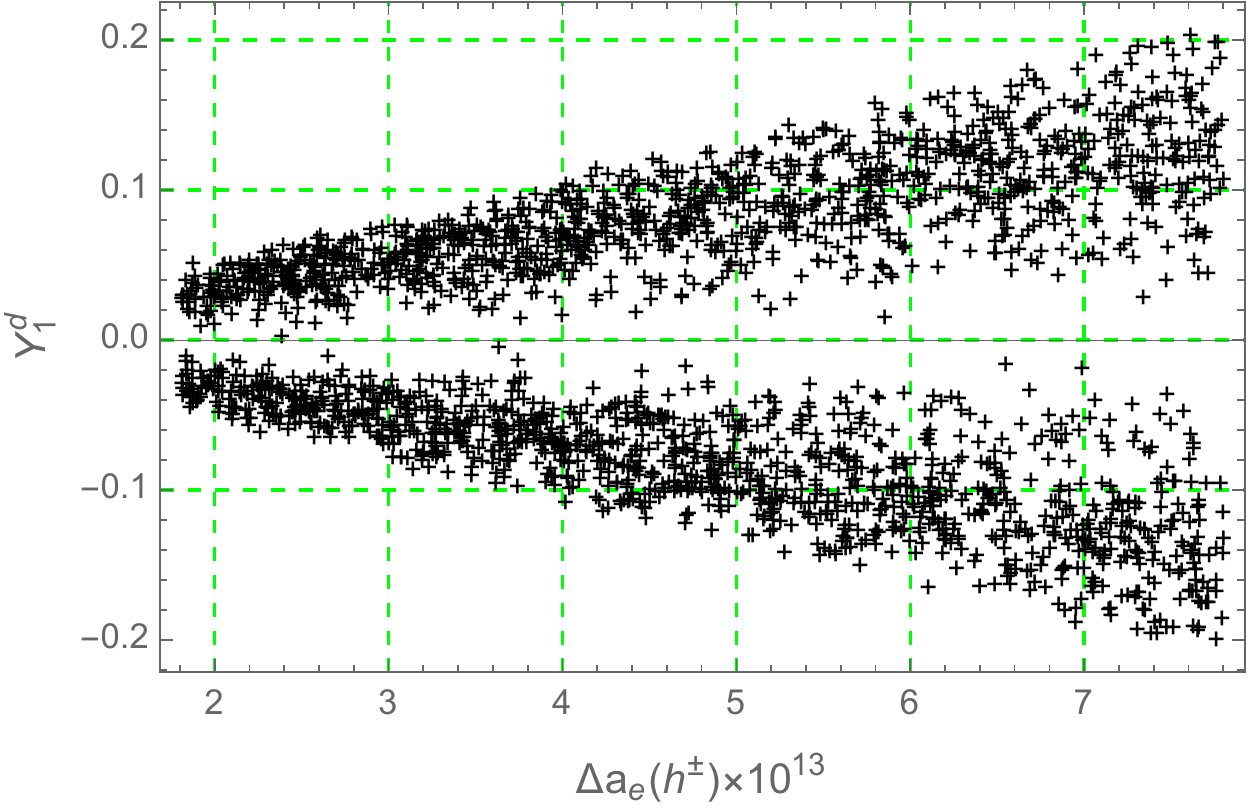}\\
	\end{tabular}
	\caption{ $\frac{Y^d_{1}}{Y^d_{d2}}$ (left panel) and $Y^d_{1}$ (right panel) as functions  of $\Delta a_{e}$ with  $f_H=t_{\beta}^{-1}$.  }\label{fig_Yh612}
\end{figure}

 As we mentioned above, the dominant contributions to $\Delta a_{e_a}$ are $\Delta a_{e_a,0}$ given in Eq. \eqref{eq_Hpm0}. This property can be seen in Fig. \ref{fig_Yh612}, showing the dependence of    the ratio  $Y^d_{1}/Y^d_{2}$ and $Y^d_{1,2}$ on $\Delta a_{e}$. 
The allowed region of this ratio $Y^d_{1}/Y^d_{2}\sim \Delta a_{e,0}(h^\pm)/\Delta a_{\mu,0}(h^\pm)$ also linearly depends on $\Delta a_e(h^\pm)$. The vertical width of the allowed region is controlled by both $1\sigma$ ranges of $\Delta a_e(h^\pm)$ and $\Delta a_{\mu}(h^\pm)$. Also, the right panel shows the linear dependence of the allowed region of $Y^d_{1}$  on $\Delta a_e(h^\pm)$.  The linear behavior of $Y^d_{2}$ is less clear than the one of $Y^d_{1}$  because $Y^d_{2}$ is also affected by the perturbative condition.

 An interesting property is that excepting $t_{\beta}$, all other parameters like $M_0$, the non-unitary parameter $\hat{x}_{\nu3}$, $Y^{d}_{1,2}$, and $s_{\varphi}$ must have lower bounds. The allowed ranges of heavy neutrino masses $M_0\ge \mathcal{O}(100)$ GeV might be confirmed by recent experimental searches in colliders such as LHC and ILC \cite{Das:2012ze, Das:2014jxa, Das:2015toa, Das:2016hof, Das:2018usr}. Because of the sizeable mixing angle   $\sim \sqrt{x_{\nu,3}}$ between ISS  and active neutrinos $\nu_{aL}$, the main production channel of heavy neutrinos $n_I$ (I=4,...,9) with mass $M_0$ at the LHC is via the Drell Yan anihilation process $u\bar{d}\to n_Ie^+_a$ mediated by the $W$ gauge boson in the $s$ channel. 
 Then the decay channel of $n_I$ can be  $n_I\to e_a^-W^+,n_{a} Z,\; n_a h$, where $h$ is the standard model-like Higgs boson. The ILC can produce heavy neutrino in the processes $e^+e^-\to \bar{n}_a n_I$ through the exchange of virtual $W$ and $Z$ bosons in the $t$ and $s$-channels, respectively. The model under consideration also predicts the production channel of a heavy neutrino pair  $e^+e^-\to \bar{n}_I n_I$ through the virtual exchange of $h^\pm_k$. In addition, the
 singly charged Higgs bosons in the model under consideration can be searched in a proton-proton collider through the processes $pp\to\gamma^*/Z^*\to h^+_kh^-_l\to (e^+_cn_{a})(\bar{n}_be^-_d)$ (with $a,b,c,d=1,2,3$), where the Yukawa couplings $Y^h$ give an important  
 contribution \cite{Calle:2021tez}. The ILC can produce two singly charged Higgs bosons $e^+e^-\to h^+_kh^-_l$ through the $n_I$ exchange in the $t$-channel. Studying these processes are beyond the scope of this work, but will be investigated in more detail elsewhere. 
 Because of the non-vanishing mixing between $h^\pm$ and the singly charged components of the Higgs doublets, another decay into a CP-odd neutral Higgs boson $A$,  such as  $h^\pm_k \to AW^\pm$, can occur \cite{Rose:2021cav}. 

\begin{figure}[ht]
	\centering\begin{tabular}{cc}
	\includegraphics[width=7.5cm]{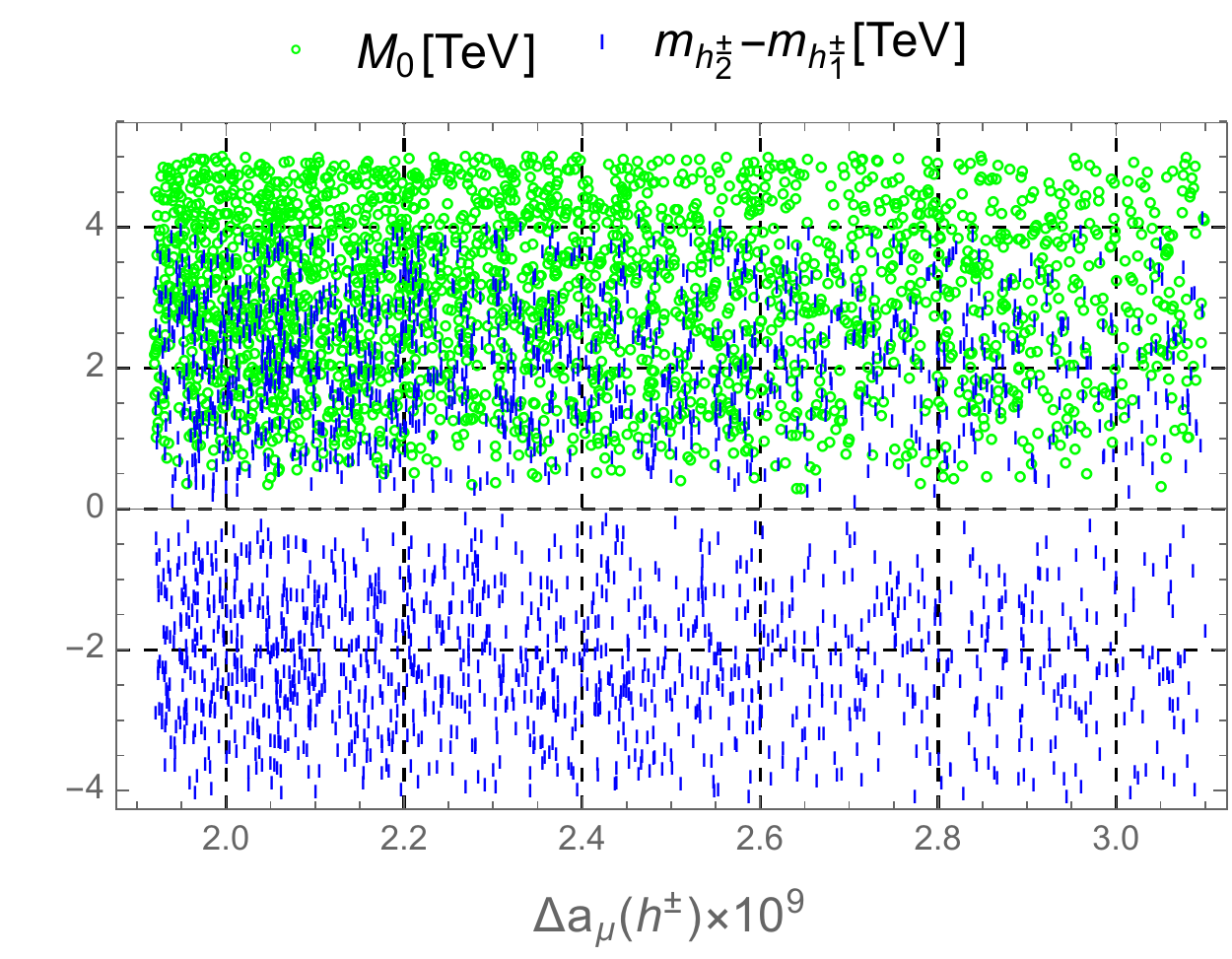}
		&	\includegraphics[width=7.5cm]{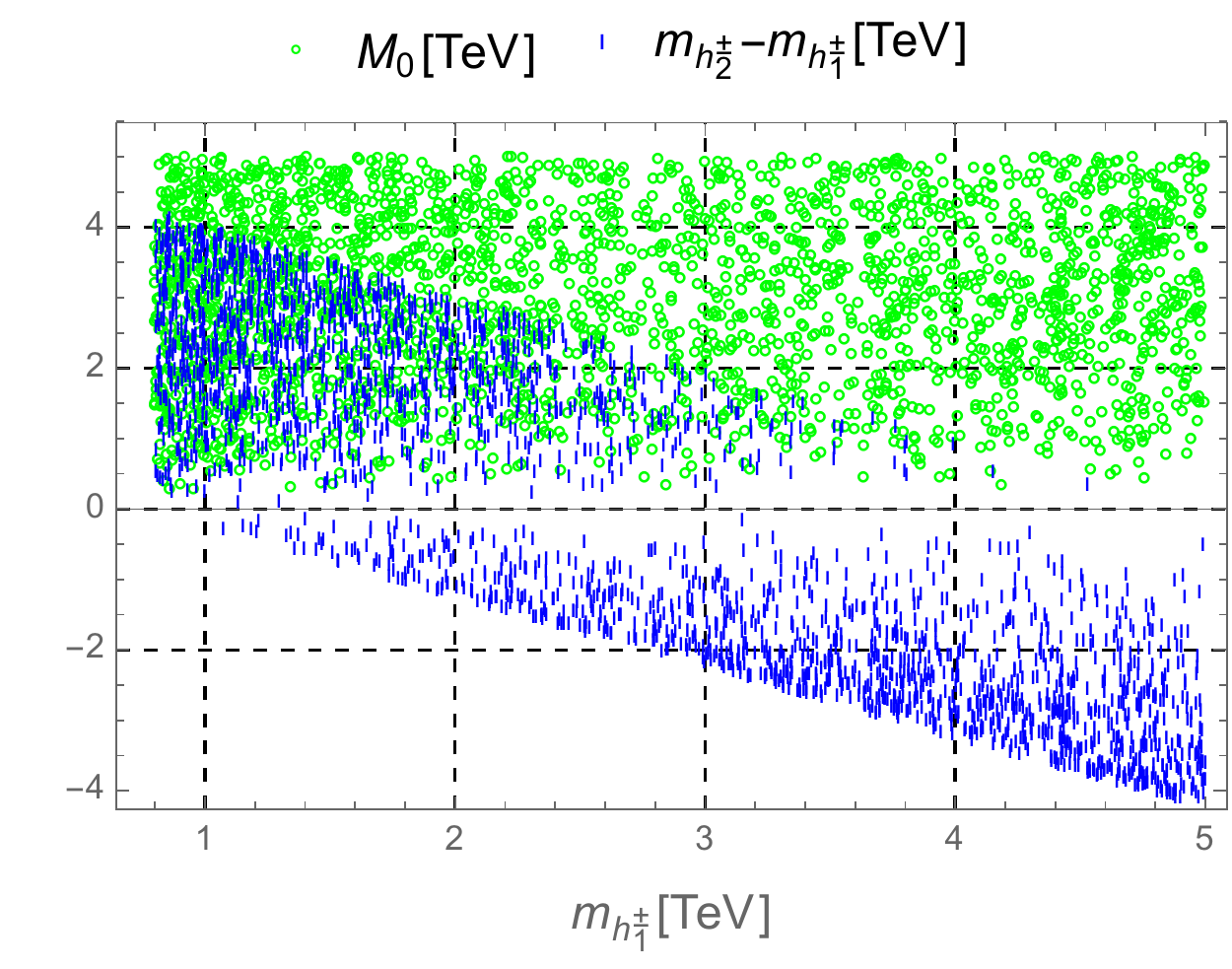}\\
	\end{tabular}
	\caption{ The correlations of $ \Delta a_{\mu}(h^\pm)$ and  $ m_{h^\pm_1}$ vs. masses $M_0$ and  $m_{h^\pm_{1,2}}$ with $f_H=t_{\beta}^{-1}$. }\label{fig_mX}
\end{figure}

The correlations relating the masses with $\Delta a_{\mu}(h^\pm)$ are shown in Fig.~\ref{fig_mX}.
We can see that $m^2_{h^\pm_1}$  must be different than $m^2_{h^\pm_2}$ and our numerical analysis indicates $\left| m^2_{h^\pm_2}-m^2_{h^\pm_1}\right|\ge 100$ GeV. The allowed regions of large $m_{h^\pm_{1,2}}$ at TeV scale  corresponding to $(g-2)_{\mu}$ experimental data may be tested indirectly through the process $\mu^+\mu^-\to\gamma^*\to h\gamma$ at multi-TeV muon colliders \cite{Yin:2020afe}.

\begin{figure}[ht]
	\centering\begin{tabular}{cc}
			\includegraphics[width=8.cm]{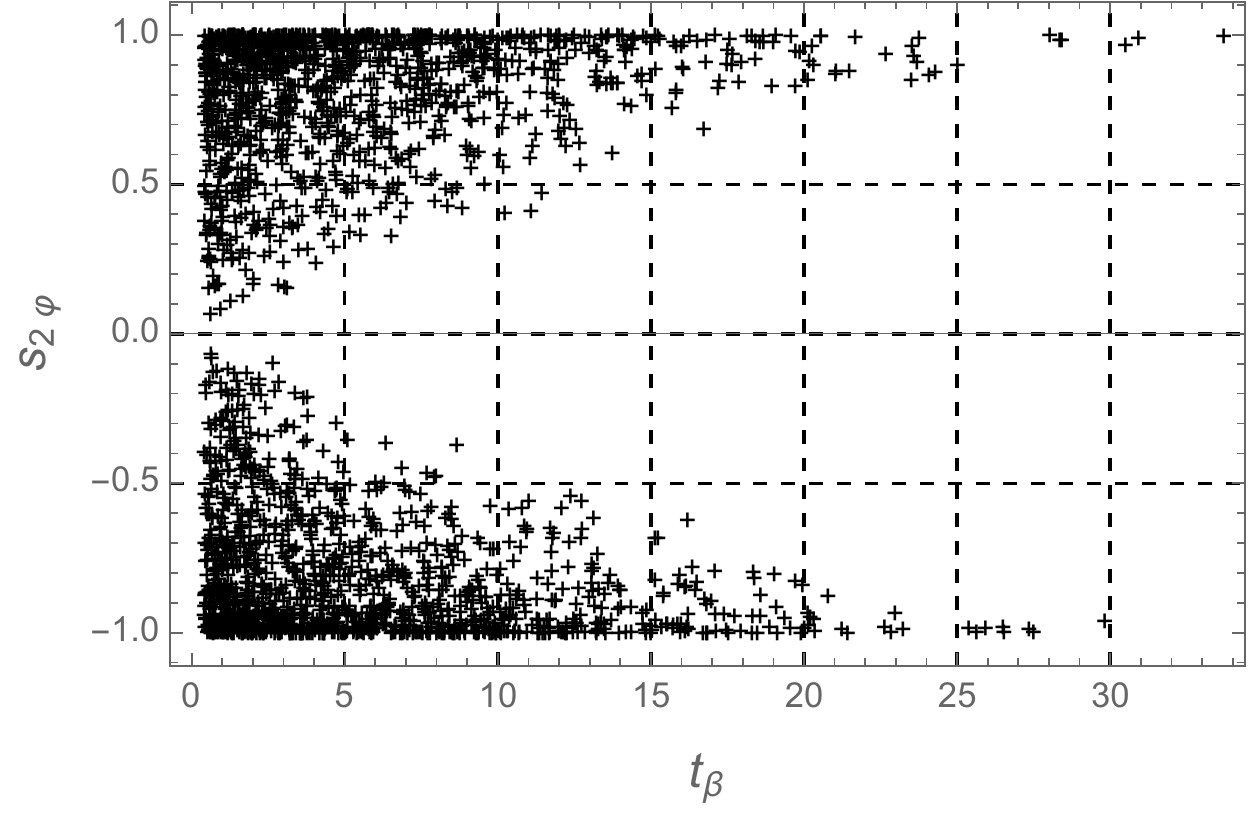}&	\includegraphics[width=8.cm]{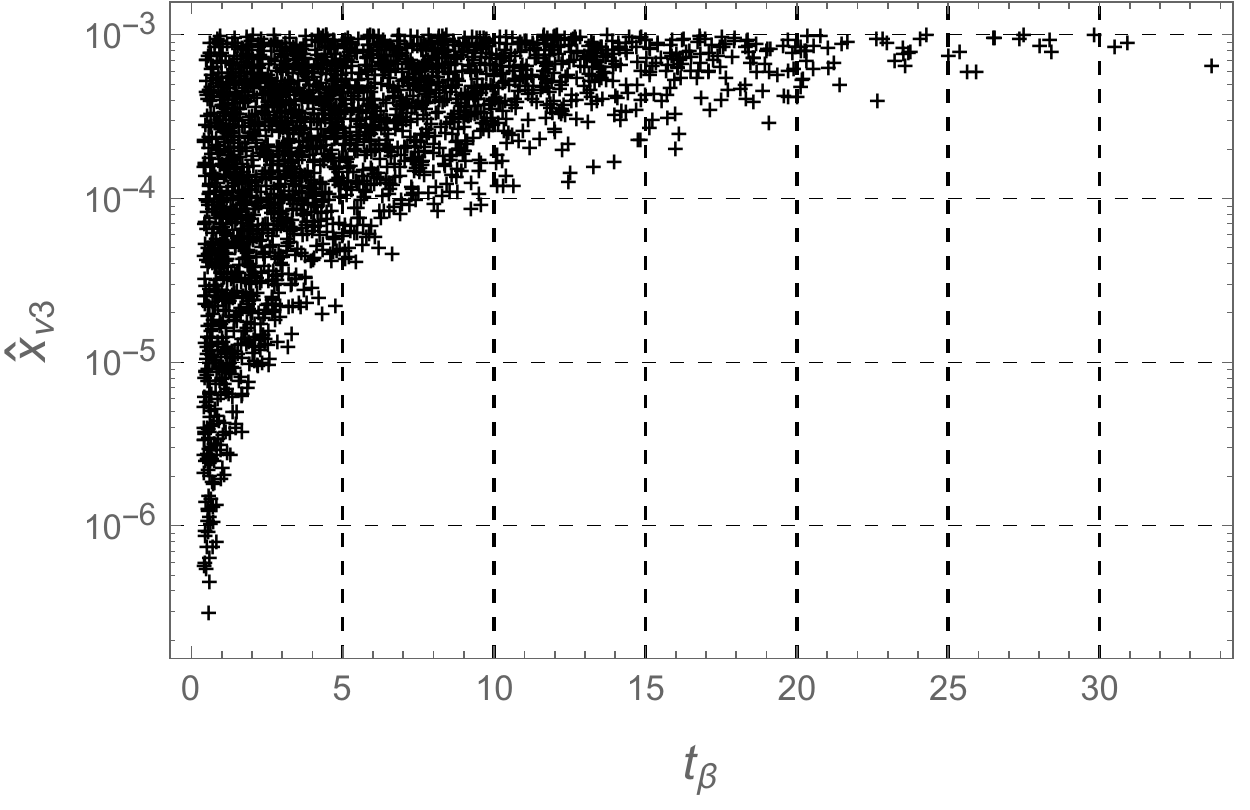}\\	
		\end{tabular}
		\caption{ The correlations of  $ t_{\beta}$ vs other free parameters  in the allowed regions with $f_H=t_{\beta}^{-1}$. }\label{fig_tbX}
\end{figure}

Finally, the  correlations showing significant dependence of free parameters and $t_{\beta}$ are given in Fig.~\ref{fig_tbX},
where the allowed regions with large $t_{\beta}\ge 20$ require both conditions of large mixing $|s_{2\varphi}|=1$ and large $\hat{x}_{\nu3}$. We obtain a small allowed range of $t_{\beta}\leq10$ that was missed in Ref.~\cite{Mondal:2021vou}. Our result is consistent with the discussion corresponding to 3-3-1 models given in Ref.~\cite{Hue:2021xap}, where the THDM is embedded.

\begin{figure}[ht]
	\centering\begin{tabular}{cc}
		\includegraphics[width=7.5cm]{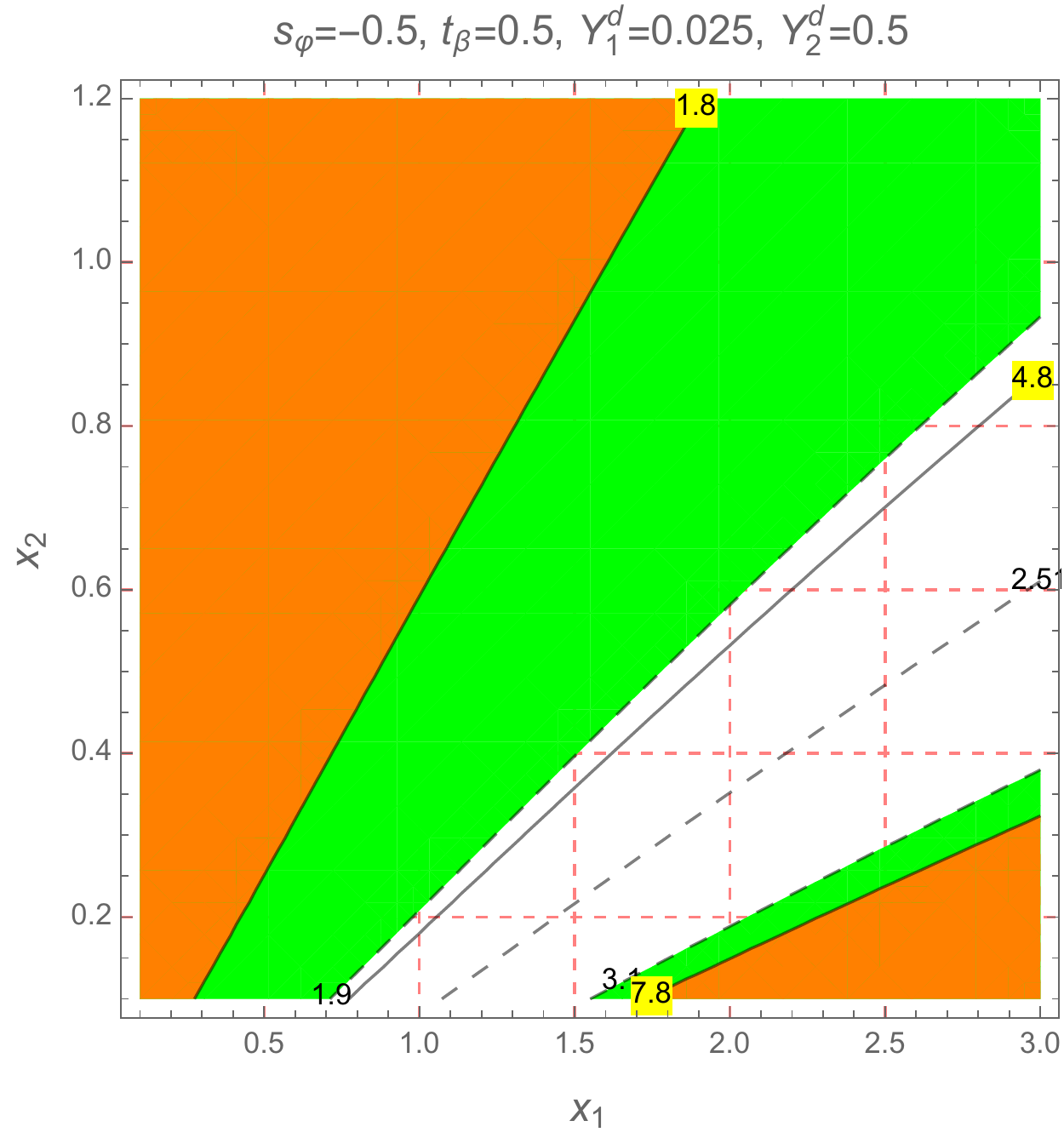}&	\includegraphics[width=7.5cm]{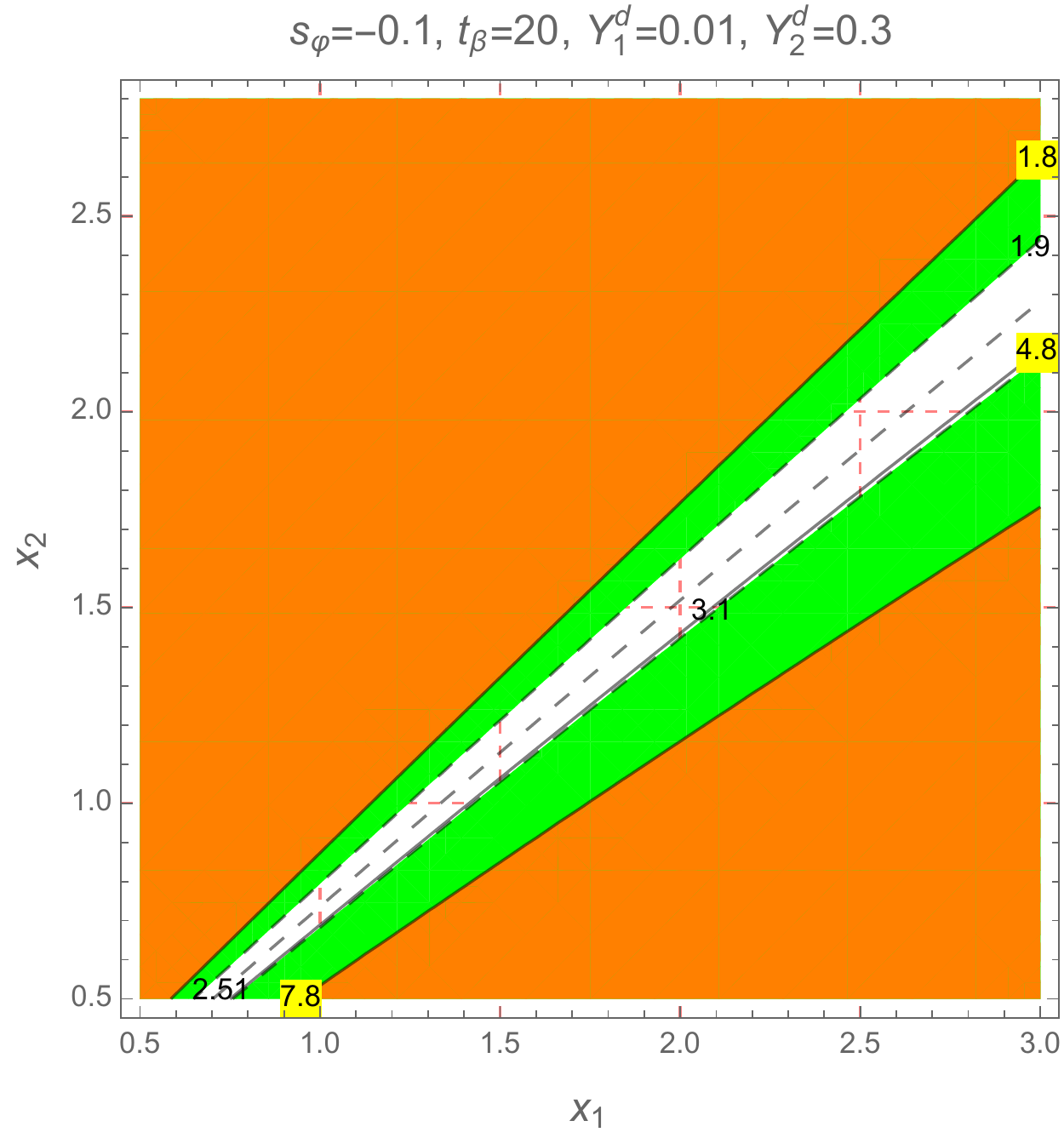}\\	
	\end{tabular}
	\caption{ Contour plots of $\Delta a_{\mu}(h^{\pm})\times 10^{9}$ and $ \Delta a_{e}(h^{\pm})\times 10^{13} $ as functions of $x_1$ and $x_2$, where $\hat{x}_{\nu3}=5\times 10^{-4}$ and $f_H=-t_{\beta}$. The green (orange) regions are excluded by the  1$\sigma$ data of $\Delta a^{\mathrm{NP}}_{\mu}$ ($\Delta a^{\mathrm{NP}}_{e}$). The black and dashed-black curves show the constant values of  $\Delta a_{e}(h^{\pm})\times 10^{13} $  and $\Delta a_{\mu}(h^{\pm})\times 10^{9}$, respectively. }\label{fig_imuae1}
\end{figure}

For the case of  $f_H=(-t_{\beta})$, i.e.,  RH neutrino and charged leptons singlets couple with the same $SU(2)_L$ Higgs doublet $\Phi_1$ in  Yukawa Lagrangian \eqref{eq_ylepton1}, for example Ref. \cite{Hue:2021xap}. Then large $a_{e_a}(h^\pm)$ in Eq. \eqref{eq_Hpm0} support large $t_{\beta}$.  In addition, even when $a_{e_a}(h^\pm)\sim s_{2\varphi}\ne0$, the first terms in the first lines of Eq.~\eqref{eq_Hpm1} or~\eqref{eq_Hpm2} may be large enough consistent with $\Delta a^{\mathrm{NP}}_{\mu,e}$ in both sign and amplitude. But the simple assumptions  of the couplings and the total neutrino mass matrix in this work are  not enough to explain both $(g-2)_{e,\mu}$ experimental data. 
 The perturbative constraint gives an upper bound on $M_0$, namely  
\begin{align} \label{eq_tbYN0bound}
	M_0 \left| \left( \hat{x}_{\nu}^{1/2}U^{\dagger}_{\mathrm{PMNS}}\right)_{ab}\right|= \frac{v c_{\beta}}{\sqrt{2}} \left|(Y^N_1)_{(a+3)b}\right|\Rightarrow M_0< \hat{x}_{\nu,3}^{-1/2} \left|U_{\mathrm{PMNS},23}\right|^{-1}v c_{\beta}\sqrt{2\pi},
\end{align}
therefore $M_0$ may be small with small $c_{\beta}$ equivalently with large $t_{\beta}$. We always have  $M_0<1.6$ TeV  for  $t_{\beta} =0.4$ and $\hat{x}_{\nu,3}=10^{-3}$.  Although $t_{\beta} \ge 0.4$ is always kept, smaller  $\hat{x}_{\nu,3}\le 10^{-3}$ may allow large $M_0$ which also allow large $m_{h^\pm_{1,2}}$.  On the other hand, from Eq.~\eqref{eq_Hpm0}, $|a_{e_a}|\sim t_{\beta} \left(x_{\nu3} \right)^{1/2} |s_{2\varphi}Y^{d}_{a}| $, hence the allowed regions  are easily satisfied for small $t_{\beta}$ and large values of other parameters. Our numerical analysis shows that large $x_{\nu3}\ge 10^{-4}$ allows all $t_{\beta}\leq 50$ and all singly charged Higgs masses at the TeV scale. Contour plots of $\Delta a_{\mu}(h^{\pm})\times 10^{9}$ and $\Delta a_{e}(h^{\pm})\times 10^{13}$  for $\hat{x}_{\nu,3} =5\times 10^{-4}$ with small $t_{\beta}=0.5$ and large $t_{\beta}=20$ are shown in Fig.~\ref{fig_imuae1}. 

\begin{figure}[ht]
	\centering\begin{tabular}{cc}
		\includegraphics[width=8.cm]{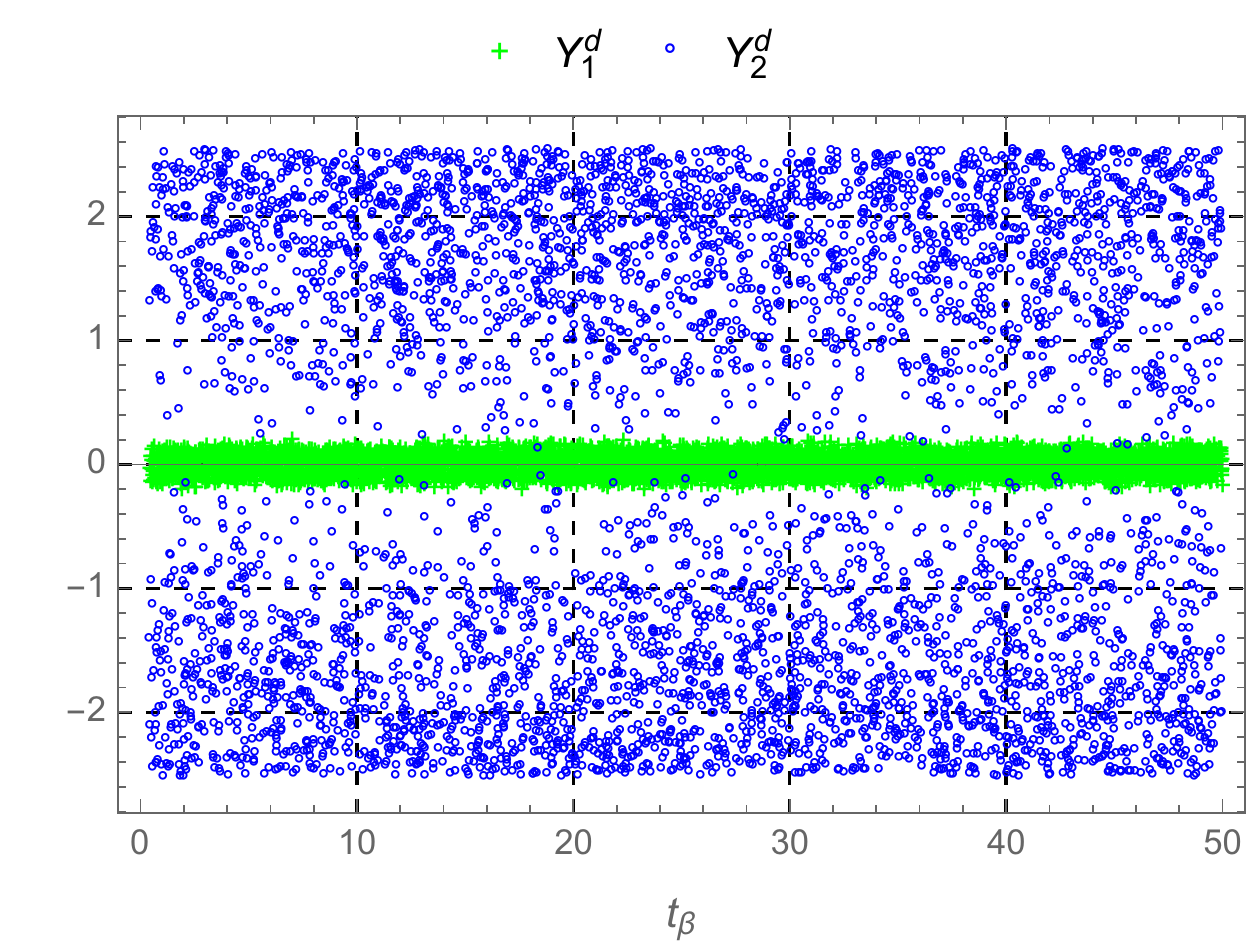}&	\includegraphics[width=8.5cm]{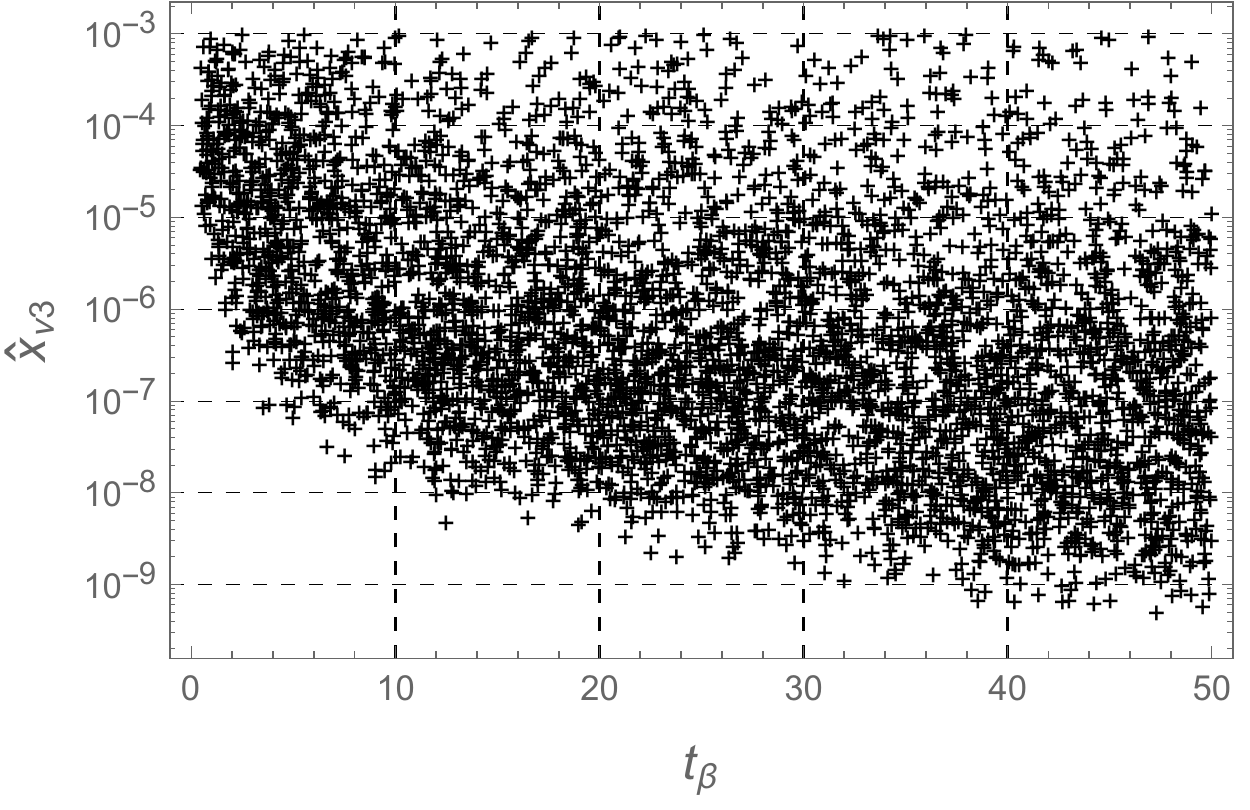}\\	
	\end{tabular}
	\caption{ The correlations of  $ t_{\beta}$ vs other free parameters  in the allowed regions with $f_H=-t_{\beta}$. }\label{fig_tbX2}
\end{figure}

Regarding the numerical analysis in the scanning range \eqref{eq_scanRanges} with $f_H=-t_\beta$, the allowed ranges are tighter than the scanning regions as shown below 
\begin{align}
	\label{eq_allowedtb}
|s_{\varphi}|>10^{-3},\;0.21\ge |Y^d_{1}|>0.002,\;   	 2.6> |Y^d_{2}| >0.02,\;  \hat{x}_{\nu3}> 3 \times 10^{-10}. 
\end{align}
In addition small $M_0\leq 30$ GeV is also allowed. The numerical results of the correlations between $t_{\beta}$ vs. $Y^{d}_{1}$, $Y^{d}_{2}$, and $\hat{x}_{\nu3}$ are shown in Fig.~\ref{fig_tbX2},
where the allowed ranges of $Y^{d}_{2}$ correspond to a rather narrow allowed range of $Y^d_{1}$, see Eq. \eqref{eq_allowedtb}. This implies that the phenomenology of the singly charged Higgs boson at colliders related with these two couplings will have some certain relations that should be experimentally verified.   

\begin{figure}[ht]
 	\centering\begin{tabular}{cc}
 		\includegraphics[width=6.cm]{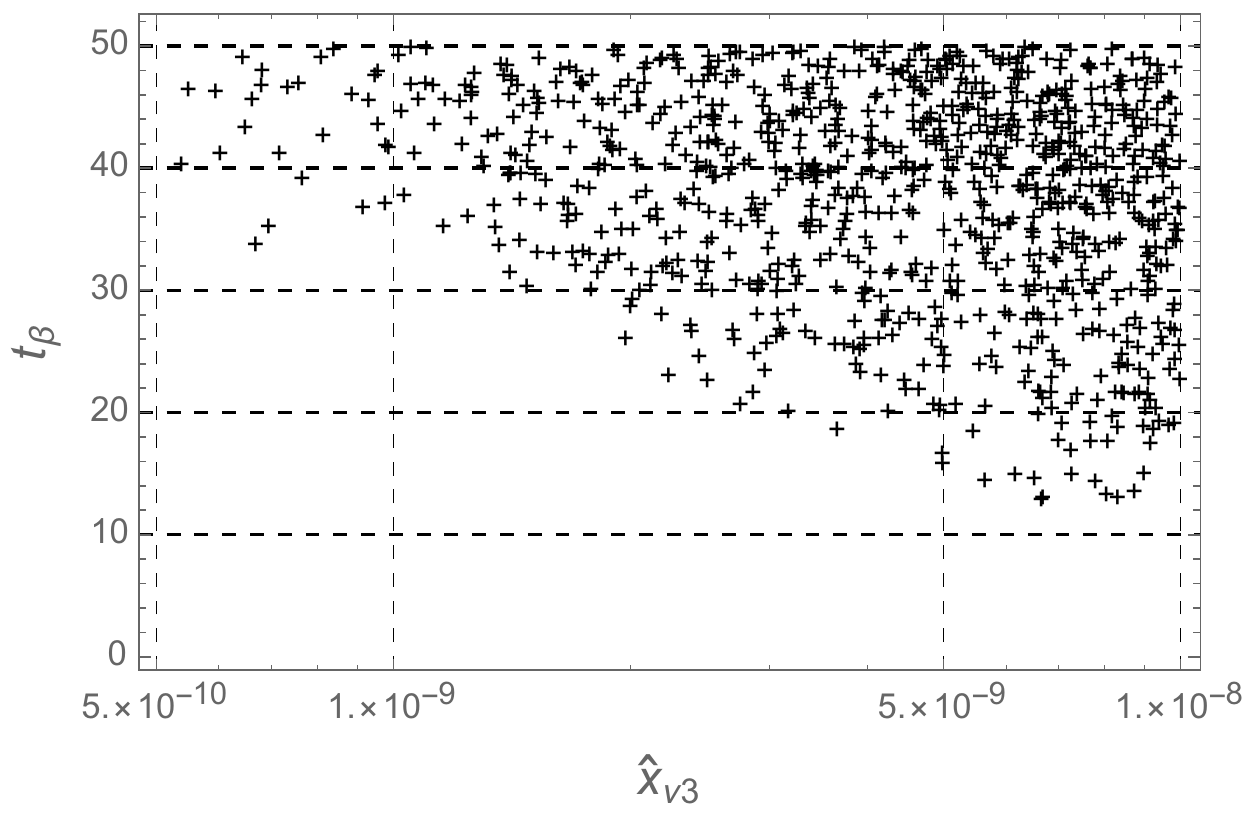}&	\includegraphics[width=6cm]{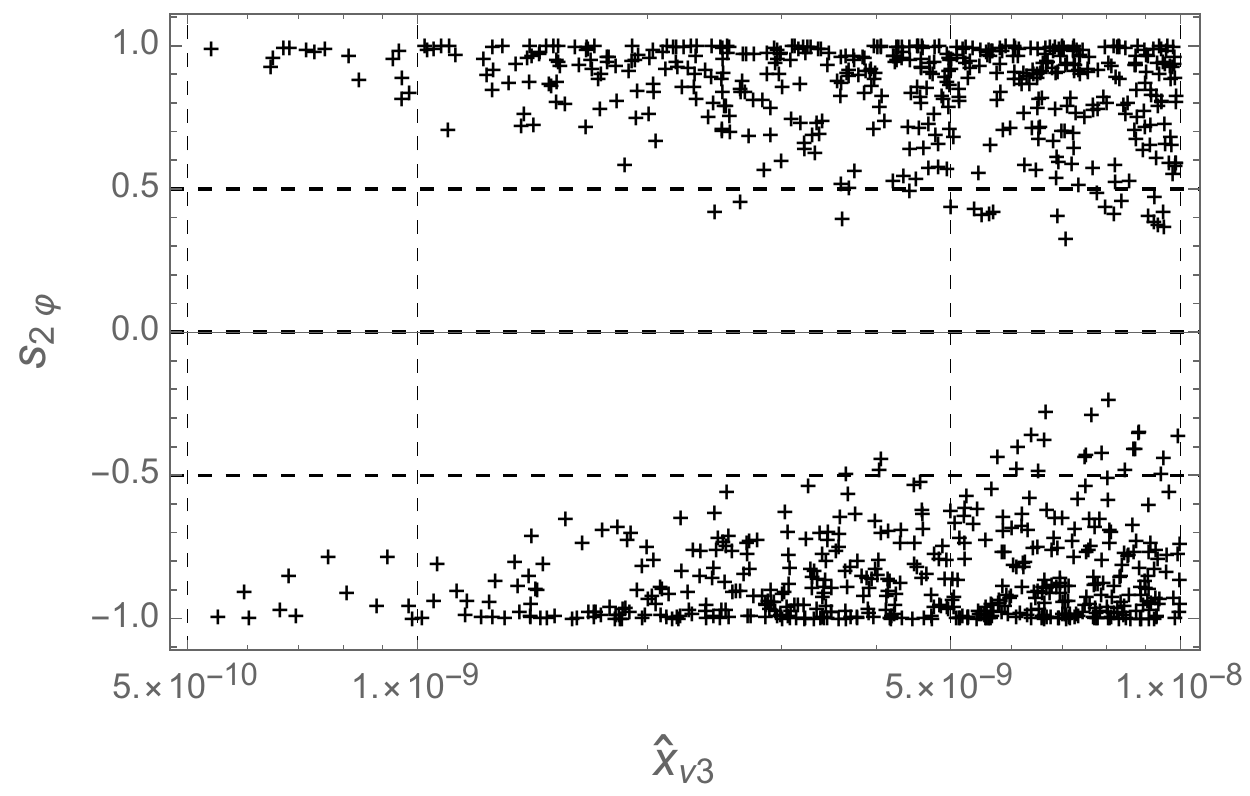}\\	
 		\includegraphics[width=6.cm]{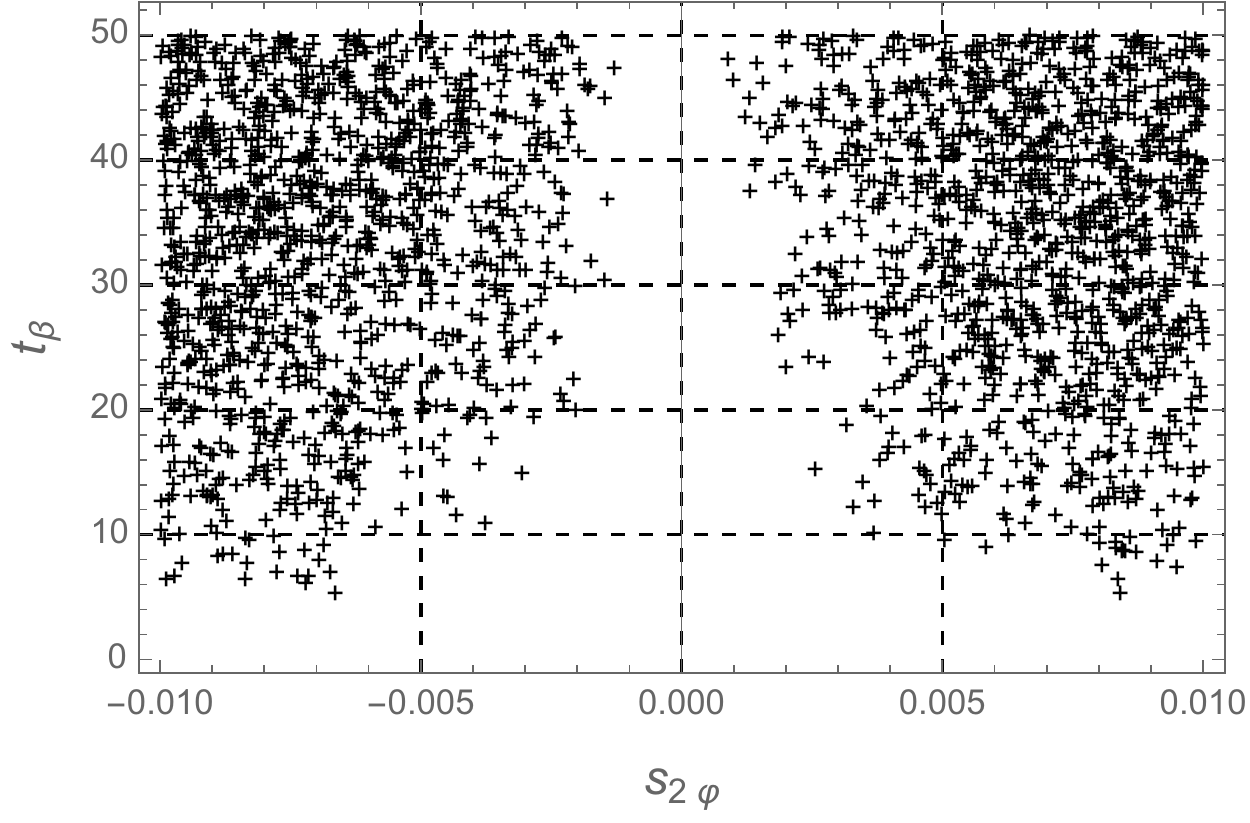}&	\includegraphics[width=6cm]{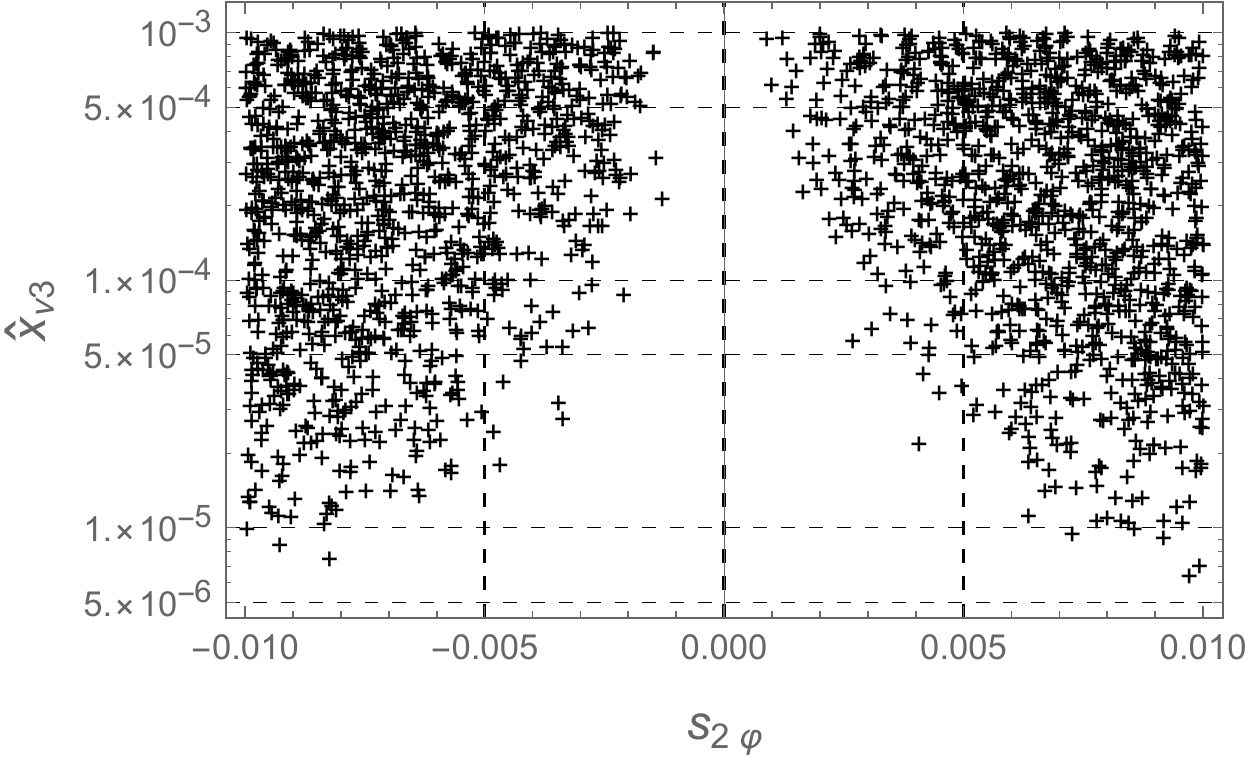}
 	\end{tabular}
 	\caption{ The correlations of  $ t_{\beta}$ vs $s_{2\varphi}$ (upper panels) and $\hat{x}_{\nu3}$ (lower panels) near lower bounds  of their allowed ranges with $f_H=-t_{\beta}$. }\label{fig_tbX2a}
 \end{figure}

 We have checked numerically that although lower bounds for allowed ranges of $s_{\varphi}$ and $\hat{x}_{\nu3}$ are tiny, but never vanishes. In addition, small allowed values of $\hat{x}_{\nu3}$ near lower bounds   require both large $t_{\beta}\to 50$ and $|s_{2\varphi}|=2|c_{\varphi}s_{\varphi}|\to1$, implying the existence of $ \Delta a_{e_{a,0}}(h^\pm)$.   Also, small allowed values of $|s_{2\varphi}|$   near the lower bound  require both large $\hat{x}_{\nu3}\to 10^{-3}$ and $t_{\beta}$. Illustrations for these comments are shown in Fig. \ref{fig_tbX2a}.
 We see also that, the relation shown in Fig. \ref{fig_Yh612} is also true with  $f_H=-t_{\beta}$.  

Finally, we have comments regarding the $(g-2)_e$ data~\cite{Parker:2018vye}: $\Delta a^{\mathrm{NP}}_e=- (8.7\pm3.6) \times 10^{-13}$. The allowed regions of parameter space corresponding to this data can be derived from the above described numerical analysis. Namely, excepting $Y^{d}_{1}$ all allowed ranges of free parameters are kept unchanged to guarantee consistency with the experimental $(g-2)_{\mu}$ data. On the other hand, $Y^{d}_{1}$ is changed into new values such that  $Y^{d}_{1}\to \frac{(-8.7\pm 3.6)}{ (4.8\pm3)}\times Y^{d}_{2}$ and exclude too large values violating the perturbative limit.  The two models under consideration predict  allowed regions of parameter space that are different  from those  discussed in Ref. \cite{DelleRose:2020oaa, Rose:2021cav} for  the THDM model where fermions couple to two Higgs doublets with the aligned assumption of the two respective Yukawa couplings $Y'_{2f}=\zeta_{f}Y'_{1f}$ with $f=u,d,\ell,\nu$, where $\nu$ denote  new right handed neutrinos $\nu_{aR}$. The large contributions to $(g-2)_{\mu}$ comes from the two-loop Barr-Zee contributions with the necessary condition of  very light neutral CP-odd Higgs boson mass, $m_A\leq 60$ GeV. On the other hand, large and negative sign of $(g-2)_e$ satisfying the $1\sigma$ experimental data comes from one-loop contribution with the condition that $\zeta_\ell \zeta_{\nu}<0$.  The model in Ref.~\cite{DelleRose:2020oaa} does not include the case $Y'_{1\nu}=0$ and $Y'_{2\nu}\neq 0$, corresponding to  $f_H=t_{\beta}^{-1}$ mentioned in our work. In addition, the region of parameter space with $\zeta_{\nu}=\zeta_{\ell}=0$ corresponding to $f_H=-t_{\beta}$ is excluded by the $1\sigma$ range of $(g-2)_e$ data. In contrast, our models always assume that $Y'_{1u}=\zeta_{\ell}=0$, and  only one of the two Yukawa coupling matrices $Y'_{1\nu}$ or $Y'_{2\nu}$ being non-zero. The Yukawa couplings $Y^d_{a}$ between only gauge singlets   $N_{a(b+3)}e_{aR}h^+$ give main one-loop contributions to both $\Delta a_{e,\mu}$, leading to nearly linear relations of these two quantities. Finally, our models predict regions of parameter space that successfully accommodate the  
experimental data on both $(g-2)_{e,\mu}$ anomalies without the requirement of small $m_A$ and rather light $m_{h^\pm_k}\sim \mathcal{O}(10^{2})$ GeV. Therefore, our models will be a another solution for the $(g-2)_{e,\mu}$ anomalies if a light CP odd scalar $A$  
is excluded by future experiments.   

\section{ \label{eq_conclusion} Conclusion}
In this work we have shown that the appearance of heavy ISS neutrinos and singly charged Higgs bosons  is a very promoting  solution to explain the experimental data on both $(g-2)_{\mu,e}$ anomalies in  many different types of THDM, and in regions of parameter space allowing heavy singly charged Higgs boson masses up to the TeV scale and small values of the $\tan\beta$ parameter satisfying $\tan\beta\geq 0.4$. In particularly, the  most important terms $\Delta a_{e_a,0}$ given in Eq. \eqref{eq_Hpm0} are enough to explain successfully the experimental AMM data of both $a_{e}$ and $a_{\mu}$ in THDM, including the model type I, where  other  loop contributions to $\Delta a_{e_a}$ caused by power of factor $t^{-1}_{\beta}$ are  suppressed. In other types of THDM  needing large $t_{\beta}$ giving sizeable loop contributions to $\Delta a_{e_a}$,  light masses of new Higgs bosons around few hundred GeV in the loop are also necessary to successfully accommodate the recent experimental AMM data. The presence of $\Delta a_{e_a,0}$ is an alternative way to explain the AMM data if light mass ranges are excluded by future collider searches. This solution will enlarge the allowed regions of parameter spaces of the THDM which can simultaneously explain the $(g-2)_{e,\mu}$ data thanks to the one loop exchange of electrically charged Higgs bosons with masses within the LHC reach. The existence of $\Delta a_{e_a,0}$ also yieds the following consequences: i) the non-zero mixing $s_{2\varphi}\neq0$, ii) the non-unitary parameter $\hat{x}_{\nu,3}$ has a lower bound $\hat{x}_{\nu,3}\ge 4\times 10^{-7}$ for $f_H= t_\beta^{-1}$ and $\hat{x}_{\nu,3}\ge \mathcal{O}(10^{-10})$ for $f_H=-t_{\beta}$,  iii) and lower bounds of new heavy neutrino masses are of the order of $\mathcal{O}(100)$ GeV. Tiny but non vanishing values of $\hat{x}_{\nu,3}\sim  10^{-10}$ 
require very large $t_{\beta}$ and $|s_{2 \varphi}|\to1$. Despite the large number of parameters, the model is economical with a small amount of BSM fields, much lower than the corresponding to several models considered in the literature. Besides that, apart from explaining the $(g-2)$ anomalies, the model considered in this paper can feature interesting collider signatures mainly related with electrically charged scalar and heavy neutrino production at the LHC, that can be useful to test that theory at colliders.

\section*{Acknowledgments}
We thank Prof.  Ray Volkas, Dr. Claudio Andrea, Dr. Lei Wang, and Dr. Wen Yin  for useful comments. L. T. Hue is thankful to  Van Lang University. This research is funded by the Vietnam National Foundation for Science and Technology Development (NAFOSTED) under the grant number 103.01-2019.387 as well as by ANID-Chile FONDECYT 1210378, ANID PIA/APOYO AFB180002,  and
Milenio-ANID-ICN2019\_044.

\end{document}